%% file: ArxivV2.tex
\documentclass[journal]{IEEEtran}
\usepackage{amsmath,amssymb,amscd,latexsym,dsfont}
\usepackage{comment}
\usepackage{color}
\usepackage{subfigure}
\usepackage{enumerate}
\usepackage{stfloats}
\usepackage{psfrag}
\usepackage{setspace}
\usepackage{cite}
\usepackage{balance}
\usepackage{enumitem}
\usepackage{tikz}
\usepackage{pgfplots,amsfonts}
\usepackage{pgfplotstable}
\usetikzlibrary{shapes}
\usetikzlibrary{spy}
\usetikzlibrary{fit}
\usetikzlibrary{shapes.multipart}
\usetikzlibrary{positioning}
\input{myFigures.tex}

\input{myStyles.tex}
\input{myMacros.tex}

\input{myPackages.tex}

\IEEEoverridecommandlockouts

\title{Polarization-ring-switching for nonlinearity-tolerant geometrically-shaped four-dimensional formats maximizing generalized mutual information
}
\author{Bin Chen,~\IEEEmembership{Member,~IEEE}, Chigo Okonkwo,~\IEEEmembership{Senior Member,~IEEE}, Hartmut Hafermann, \\ and Alex Alvarado,~\IEEEmembership{Senior Member,~IEEE}. 
\thanks{B. Chen is with  the School of Computer and Information, Hefei University of Technology (HFUT), Hefei, China. B. Chen is also with Eindhoven University of Technology, Eindhoven 5600 MB, The Netherlands.  (e-mail: b.c.chen@tue.nl)}
\thanks{C. Okonkwo  is with the Institute for Photonic Integration, Eindhoven University of Technology, Eindhoven 5600 MB, The Netherlands (e-mail: c.m.okonkwo@tue.nl).}
\thanks{A. Alvarado is with the Information and Communication Theory Lab, Signal Processing Systems Group, Department of Electrical Engineering, Eindhoven University of Technology, 5600 MB, Eindhoven, The Netherlands (e-mail:~a.alvarado@tue.nl).}
\thanks{Hartmut Hafermann is with the Optical Communication Technology Lab, Paris Research Center, Huawei Technologies France SASU, 92100 Boulogne-Billancourt, France (e-mail:~hartmut.hafermann@huawei.com).}
\thanks{Research supported by Huawei France through the NLCAP project.}
\thanks{Thw work of B. Chen is supported  by the National Natural Science Foundation
of China (NSFC) under Grant 61701155. The work of A. Alvarado is supported by the Netherlands Organisation for Scientific Research (NWO) via the VIDI Grant ICONIC (project number 15685).}
}

\begin{document}
\maketitle

\begin{abstract}
In this paper, a new four-dimensional 64-ary polarization ring switching  (4D-64PRS) modulation format with a spectral efficiency of $6$~bit/4D-sym is introduced. The format is designed by maximizing the generalized mutual information (GMI) and by imposing a constant-modulus on the 4D structure. The proposed format yields an improved performance with respect to state-of-the-art geometrically shaped modulation formats for bit-interleaved coded modulation systems at the same spectral efficiency.
Unlike previously published results, the coordinates of the constellation points and the binary labeling of the constellation are \emph{jointly} optimized. 
When compared {with} 
polarization-multiplexed 8-ary quadrature-amplitude modulation (PM-8QAM), gains of up to $0.7$~dB in signal-to-noise ratio are observed in the additive white Gaussian noise (AWGN) channel. For a long-haul nonlinear optical fiber system of $8,000$~km, gains of up to $0.27$~bit/4D-sym ($5.5$\% data capacity increase) are observed. These gains translate into a reach increase of approximately $16$\% ($1,100$~km). The proposed modulation format is also shown to be more tolerant to nonlinearities than PM-8QAM. Results with LDPC codes are also presented, which confirm the gains predicted by the GMI.
\end{abstract}



\section{Introduction, Motivation and State of the Art}\label{Sec:Introduction}

\IEEEPARstart{S}{ignal} shaping can be used in optical fiber communications to  close the gap to the channel capacity, either via 
probabilistic shaping 
(PS) or 
geometric shaping
(GS). In the former, long coded sequences induce certain nonuniform probability distribution on the constellation points \cite{Buchali2016,TobiasJLT16,BochererECOC2017,Buchali2017,Maher2017}. This can be achieved, e.g., by using trellis shaping \cite{trellisshaping}, shell mapping \cite{lang1989,laroia1994}, or distribution matching (DM) techniques \cite{dyadic}. In geometrical shaping, non-equidistant constellation points are used with the same probabilities. Recently, geometrical shaping has received considerable attention in the optical communications literature \cite{Qu2017,ZhangECOC2017,Kojima2017JLT,Millar2018_OFC,BinECOC2018,BinICTON2018}. For the additive white Gaussian noise (AWGN) channel, both PS and GS achieve Shannon's channel capacity when the number of constellation points tends to infinity. Hybrid approaches have been investigated in \cite{8346117,Cai17OFC}.

{
Among a number of probabilistic shaping approaches, PS  based on constant composition
distribution matching (CCDM) is particularly
promising  in optical fiber communications since it  is compatible with conventional quadrature amplitude modulation (QAM) and   offers near-optimal linear shaping gain.
An additional benefit of PS is that it enables rate adaptivity by changing the probability distribution with very fine granularity.
Although PS has superior achievable information rates (AIR) performance for a finite number of constellation points with respect to GS \cite[Sec.~4.2]{BICM_book}, \cite{Steiner2017}, PS requires the use of  high-precision arithmetic in implementation. 
Distribution matching (DM) encoders are typically used in PS and they offer very good performance for long block lengths. However, DM encoders for long block lengths are difficult to realize in practice and hardware implementation of the DM remains a
significant challenge \cite{BochererECOC2017}. Furthermore, for short block lengths, DM encoders suffer from rate losses, which limit the DM implementation in parallel to reduce  processing latencies.
Research is therefore now dedicated to find
improved DM architectures for short blocklengths, e.g.,  shell mapping \cite{SchulteWCL2019}, enumerative sphere shaping \cite{GultekinISIT2018,GultekinarXiv2019} and partition-based DM \cite{2018Tobias_PBDM}.}

{
The main drawbacks of GS are tighter requirement on the resolution of the digital-to-analog and analog-to-digital converters and the need for a modified demapper and equalizer.  For multidimensional modulation formats, GS also increases the computational complexity of the demapper,  as in this case, Euclidean distances for all \emph{multidimensional} symbols need to be calculated.
Several studies proposed low-complexity soft-demapping solutions as good trade-offs between performance
and complexity \cite{YoshidaECOC2016,BendimeradECOC2018,NakamuraJLT2018}.
Just like PS, GS also increases the nonlinear effects of the fiber. 
Different studies have shown that these nonlinear effects can be mitigated by combining multi-dimensional modulations with GS \cite{Shiner:14,ReimerOFC2016,Kojima2017JLT,Bendimerad:18}. In this paper, we consider this approach using AIRs as a design metric.
}

AIRs such as mutual information (MI) and generalized mutual information (GMI) have emerged as practical tools to design optical fiber communication systems \cite{Alvarado2015_JLT,Schmalen17}. AIRs have also been used to design modulation formats and to predict the performance of forward error correction (FEC)\cite{AlvaradoJLT2015,AlvaradoJLT2018}. GMI can be directly connected to modern binary soft-decision forward error correction (SD-FEC) 
based on bit-interleaved coded modulation (BICM). The practical relevance of GMI makes it the preferred alternative for optical fiber communication systems design. On the other hand, the MI can be used as an upper bound on the performance of binary SD-FEC performance (when combined with a nonbinary modulation schemes).
However,  the multidimensional
modulations optimized for  MI  have been shown to be non-optimal in terms of GMI
\cite{Alvarado2015_JLT}. Hence, using the GMI as a figure of merit is essential to design a modulation, which is well-suited for bit-wise decoders.

As mentioned above, both PS and GS typically increases the (modulation-dependent) nonlinear effects in the fiber. In particular, in multi-span long-haul transmission systems, nonlinear interference noise (NLIN) modelling and numerical simulations have shown that both PS and GS decrease the effective signal-to-noise ratio (SNR) due to the increasing contribution of modulation format dependent NLIN. At the same time, PS and GS increase the AIRs. If the shaping is properly designed, the effective SNR decrease is less than the AIR increase, and therefore, an overall performance improvement is observed \cite{TobiasJLT16,RasmusECOC2018}.

In order to cover a wide range of channel conditions in flexible networks with multiple modulation formats for efficient network usage, multiple GS-based modulation formats with different spectral
efficiency  have been investigated in the optical communications literature. This includes for example GS-16QAM \cite{Qu2017,BinECOC2018}, GS-32QAM \cite{ZhangECOC2017}, GS-64QAM\cite{BinECOC2018,BinICTON2018}, GS-256QAM\cite{BinECOC2018} and GS-APSK \cite{Kojima2017JLT,Millar2018_OFC}. 
GS has been reported to outperform PS constellations in numerical simulations \cite{Millar2018_OFC}, which is  particularly noticeable for dispersion managed links \cite{Kojima2017JLT,Millar2018_OFC}.

PM-8QAM with 3 bit/2D-sym  plays an important role in filling the gap between quaternary phase shift keying (QPSK) and 16QAM in terms of bit rates and reach.
To achieve the same spectral efficiency (3 bit/2D-sym) with improved performance, 
 advanced 2D modulation formats, such as 
 Circular-8QAM\cite{MuellerOFC2015} and optimized-8QAM\cite{ZhangECOC2015}, have been reported to outperform regular PM-8QAM. 
As an alternative, multidimensional (MD) modulation formats  have also attracted a lot of attention  in optical communications by transmitting information jointly in each
degree of freedom (polarization, time slot, wavelength,
modes, spatial channels, etc.).
The most popular 4-dimensional (4D) modulation formats were proposed to exploit the natural 4 dimensions in the optical field: in-phase and quadrature from two orthogonal polarizations. In this paper, we target this 4D design approach for the important case of constellations with 6 bit/4D-sym. 

Recently, three 4D modulation formats 4D-64SP-12QAM \cite{NakamuraECOC2015}, 4D-8QAM \cite{ReimerOFC2016} and 4D-2A8PSK \cite{Kojima2017JLT} have been shown to be superior to PM-8QAM and other previous modulation formats for BICM system.
These 4D modulation formats are designed  by either increasing the Euclidean distances \cite{NakamuraECOC2015} and minimizing the uncoded bit error rate (BER) \cite{ReimerOFC2016} in 4D space (improved linear performance), or by increasing the tolerance to nonlinearity (improved nonlinear performance) \cite{Kojima2017JLT}. 
 In \cite{NakamuraECOC2015}, 4D-64SP-12QAM was proposed, which is a subset of PM-16QAM, and where the points selection of constellation points was made to maximize Euclidean distances. 4D-64SP-12QAM outperforms regular 16QAM and also provides 0.2 dB gain over PM-8QAM (linear gain).
4D-8QAM was optimized in \cite{ReimerOFC2016} 
by minimizing the uncoded BER for the AWGN channel, providing a $0.5$~dB gain over PM-8QAM.
4D-2A8PSK by Kojima et al. \cite{Kojima2017JLT} has been shown to be superior to many other formats in both the linear and nonlinear regimes. This is due to its large Euclidean distance, a 4D constant-modulus constraint, and Gray labeling. The 4D constant-modulus is obtained by forcing the two polarization to have complementary amplitudes from two 8-ary phase shift
keying (PSK) with two different amplitudes. 
In \cite{Kojima2017JLT}, the GMI was optimized in terms of set-partitioning coding and two ring ratios only.
The obtained linear performance improvement of 4D-2A8PSK over PM-8QAM was $0.3$~dB.

From the literature review above, it is still unclear what is the largest GMI gain that can be achieved by designing a 4D modulation format. 
Since a 4D constant-modulus property is known to be effective in reducing the modulation-dependent NLIN (especially for dispersion-managed links), the focus of this paper is on  maximizing GMI performance under a constant-modulus constraint. 
We first reveal that the constant-modulus property can significantly reduce the modulation-dependent nonlinear noise through theoretical analysis of the effective SNR, which we believe also applies to 4D modulation formats. 
Secondly, we use GMI to design a 4D modulation format with a spectral efficiency of 6 bit/4D-sym for BICM system. The coordinates of the constellation points and the binary labeling of the constellation are \emph{jointly} optimized. 
Thirdly, we show that the proposed {four-dimensional 64-ary polarization ring switching  (4D-64PRS) modulation outperforms all previously known  geometric shaped modulation formats of the same spectral efficiency, in both the linear and nonlinear regimes.}  Shaping gains of 0.7~dB and 0.4~dB can be achieved with respect to PM-8QAM and 4D-2A8PSK for the AWGN channel, respectively.
Nonlinear optical channel simulations confirm that constant-modulus 4D-64PRS modulation formats increase the effective SNR with respect to that of regular QAM modulation
and that the proposed 4D modulation format can further increase AIRs. This gain includes a 0.16~dB nonlinear gain that is additional to the linear gains. These gains translate into reach increases in the range of $16$\%. 

This paper is organized as follows. In Sec.~\ref{sec:model}, the system model, the design methodology and the proposed modulation format are introduced. In Sec.~\ref{Sec:NumericalResults}, numerical results for the proposed 4D modulation format are shown after both AWGN and nonlinear optical fiber simulations. The importance of joint optimization of coordinates and labeling in 4D modulation design is discussed in Sec.~\ref{sec:Discussion}. Finally, conclusions are drawn in Sec.~\ref{Sec:Conclusions}.

\begin{figure*}[!htbp]
\centering
  \includegraphics[width=0.95\textwidth]{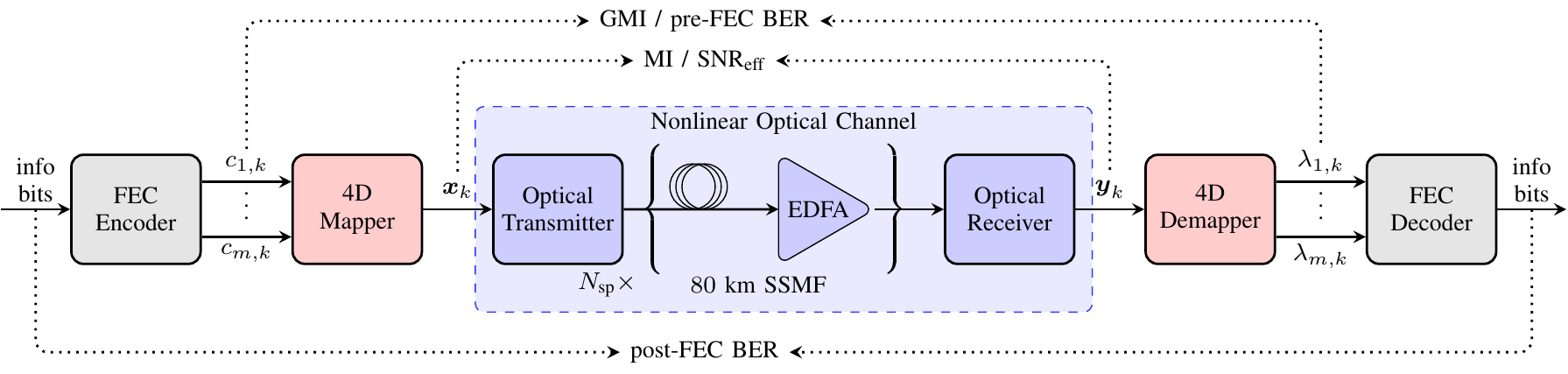}
  \vspace{-1em}\caption{System model under consideration. The nonlinear optical channel is modeled using a channel with an optical
fiber link comprising multiple spans ($N_\text{sp}$ is the number of spans). Each  span is followed by an EDFA.
  The MI  defined in \eqref{eq:MI} and GMI in \eqref{eq:GMI} respectively, are also shown.
    }
\label{fig:MI_GMI}
\end{figure*}

\section{AIR-based Geometric Shaping Optimization}\label{sec:model}

Fig.~\ref{fig:MI_GMI} shows the model under consideration, representing a generic multi-span optical fiber communication system. At the transmitter side, a FEC encoder generates $m$ binary code bits
$c_{1,k},\ldots,c_{m,k}$, where $k=1,2,\cdots, N_\text{sym}$ represents discrete time and  $N_\text{sym}$ is the number of symbols transmitted over the optical channel. These code bits are then mapped to symbols with $N$ real dimensions drawn from a constellation with $M=2^m$ constellation points.
{In coherent fiber optical  communication systems, both quadratures and both polarizations of the electromagnetic field are used (each polarization includes two real dimensions, in-phase $I$ and quadrature $Q$).
This naturally results in a four-dimensional (4D) signal space ($N = 4$).}
For a conventional 2D modulation format, i.e. 8QAM, the symbols transmitted over two polarization are statically independent of each other and can be considered as  (polarization-multiplexed) 4D symbols.
However, a true 4D symbol will be jointly modulated in the 4D space created by the two polarizations.
The resulting 4D symbols $\boldsymbol{x}_k$ are transmitted through a nonlinear optical channel. This channel consists of $N_\text{sp}$ spans of optical fibers, each of them followed by an erbium-doped fiber amplifier (EDFA).
At the receiver side, after opto-electronic conversion and digital signal processing, the symbols $\boldsymbol{y}_k$ are obtained. These symbols are then demapped into soft bits (represented as logarithmic likelihood ratios, LLRs) $\lambda_{1,k},\ldots,\lambda_{m,k}$, which are then passed to the SD-FEC decoder.\footnote{Throughout this paper we use the following notation. Matrices are denoted by blackboard bold letters $\mathbb{X}$.
Random variables are denoted by capital letters $X$ and random vectors by boldface letters $\bX$. Their corresponding realizations are denoted by $x$ and $\bx$. All vectors are row vectors, where $(\cdot)^T$ denotes transpose.
Expectations are denoted by $\mathbb{E}\left[\cdot\right]$ and the $p$-norm is defined as $||\bX||^p=|X_1|^p+|X_2|^p, \dots, |X_N|^p$. Conditional probability density
functions (PDFs) are denoted by $f_{\bY|\bX}(\bY|\bX)$.}

The largest AIR for  {an MD} memoryless channel with input $\bX$ and output $\bY$ is given by the MI: 
\begin{equation}\label{eq:MI}
I(\bX;\bY)=\mathbb{E}\left[\log_2\frac{f_{\bY|\bX}(\bY|\bX)}{f_{\bY}(\bY)}\right],
\end{equation}
where $f_{\bY|\bX}$ is the channel law. The most popular AIR for a bit interleaved coded modulation (BICM) is the GMI:
\begin{equation}\label{eq:GMI}
G=\sum_{i=1}^m I(C_i;\bY) 
= \sum_{i=1}^m\mathbb{E}\left[\log_2\frac{f_{\bY|C_i}(\bY|C_i)}{f_{\bY}(\bY)}\right],
\end{equation}
where $\bC=[C_1,C_2,\dots,C_m]$ is a random vector representing the transmitted  bits $[c_{1,k},c_{2,k},\dots,c_{m,k}]$
at time instant $k$, which  are mapped to the corresponding symbol $\bx_k$. LLRs are simply another way of representing (bit-wise) information, and thus, the GMI can also be analyzed in terms of these LLRs, as schematically shown in Fig. \ref{fig:MI_GMI}.

The transmitted symbols $\bX$ are {designed} to be 4D symbols with 4 real dimensions  drawn uniformly from a discrete constellation. The $i$th constellation point is denoted by $\bs_i=\left[s_{i,1},s_{i,2},s_{i,3},s_{i,4}\right] \in\mathbb{R}^4$ with $i=1,2,\dots,M$. We use the $M\times 4$ matrix 
$\mathbb{S}=[\bs_1^T,\bs_2^T,\ldots,\bs_M^T]^T$ to denote the 4D constellation. The $i$th constellation point $\bs_i$ is labeled by the length-$m$ binary bit sequence $\bb_{i}=[b_{i,1},\ldots,b_{i,m}]\in\{0,1\}^m$.
The binary labeling matrix is denoted by a $M\times m$ matrix $\mathbb{B}=[\bb_1^T,\bb_2^T,\ldots\bb_M^T]^T$ which contains all unique length-$m$ binary sequences. The 4D constellation and its binary labeling are fully determined by the pair of matrices $\{\mathbb{S},\mathbb{B}\}$.

Both the MI and GMI depend on the channel law. The MI  also depends on the coordinates of the constellation points $\mathbb{S}$. Thus, designing constellations to maximize the MI corresponds to finding the coordinates of $M$ points, that satisfy a power constraint. The GMI depends also on the binary labeling $\mathbb{L}$. The GMI-based optimization problem is therefore more complex than the one for MI, as the binary labeling also needs to be taken into account.
In \cite{Alvarado2015_JLT}, 4D constellations optimized for uncoded systems were shown to give high MI. Those constellations, however, were shown not to be well-suited for bit-wise decoders. 

In what follows, we describe the methodology used for the design of the proposed 4D modulation format and its binary labeling. The main approach taken is to jointly optimize the binary labeling and the coordinates of the 4D constellation points by taking into account both linear and nonlinear performance. We briefly describe the theoretical foundation for this approach. Later in the section, details of the optimization procedure are presented.

\subsection{Theoretical Background}\label{sec:TheoreticalAnalysis}
We consider a  multi-span wavelength division
multiplexing (WDM) transmission scenario where several channels are launched into a nonlinear optical fiber. The 4D noisy received signal for the channel of interest $\bY_k$ can be modeled using the classic GN model as \cite{DarISIT2014}
\begin{equation}\label{AWGN}
\bY=\bX+\bZ^{\text{NLI}}+\bZ^{\text{ASE}},
\end{equation}
where $\bX$ are the 4D symbols transmitted over the channel of interest, and $\bZ^{\text{NLI}}$ is the interference induced by the nonlinearity of the fiber. We call this effect  nonlinear interference (NLI), which includes intra-channel and inter-channel distortions. The (accumulated) amplified spontaneous emission (ASE) noise in the system is represented by $\bZ^{\text{ASE}}$. 
In \eqref{AWGN}, both the NLI and ASE are modeled as additive Gaussian noise.\footnote{This assumption is well justified for dispersion uncompensated long-haul transmission systems.} Furthermore, \eqref{AWGN} is a  memoryless channel, which is why we dropped the time index $k$.

The effective SNR (denoted by $\text{SNR}_{\text{eff}}$) represents the SNR after fiber propagation and the receiver digital signal processing (DSP)
and is defined as \cite[Eq. (4)]{Poggiolini_JLT2014},\cite[Eq. (16)]{TobiasJLT16}
\begin{equation}\label{eq:eff}
\text{SNR}_{\text{eff}}\triangleq \frac{\mathbb{E}\left[\|\bX\|^2\right]}{\mathbb{E}\left[\|\bY-\bX\|^2\right]}=\frac{\sigma_x^2}{\sigma_z^2}=\frac{\sigma_x^2}{\sigma^2_{\text{ASE}}+\sigma^2_{\text{NLI}}},
\end{equation}
where  $\sigma_x^2$ and $\sigma_z^2$ represent the transmitted power and total noise power per two real dimensions, respectively. In \eqref{eq:eff}, $\sigma^2_{\text{ASE}}$ represents the variance of the ASE noise and $\sigma^2_{\text{NLI}}$ the variance of the NLI that includes both intra- and inter-channel distortions.

The standard Gaussian Noise (GN) model \cite{Poggiolini_JLT2014} ignores dependency of nonlinear effects  on the modulation format. Other more advanced models such as the NLIN model in \cite{Dar:13} (see also \cite{SecondiniPTL2012,SerenaECOC2013,Carena:14})
allow more accurate analysis of non-conventional modulation formats, such as modulation with PS and GS.
In these NLI models, it is assumed that the symbols transmitted over two polarizations are statistically independent of each other and each polarization uses identical 2D modulation formats. This corresponds to the so-called ``polarization-multiplexed'' systems, which includes for example PM-8QAM which we study in this paper. The NLI noise term $\sigma^2_{\text{NLI}}$ for polarization multiplexed systems in \eqref{eq:eff} can be approximated as    \cite{Dar:13,Carena:14,TobiasJLT16} 
\begin{align}\label{nli}\nonumber
\sigma^2_{\text{NLI}}\approx  
&\underbrace{\eta_1\cdot P^3}_{\text{Modulation-independent}} \\ &+\underbrace{P^3\left[\eta_2\cdot  (\mu_4-2)+\eta_3\cdot  (\mu_4-2)^2+\eta_4\cdot \mu_6\right]}_{\text{Modulation-dependent}},
\end{align}
where $P$ is the optical launch launch power, $\eta_1$,  $\eta_2$, $\eta_3$ and  $\eta_4$  are constants (for a given system configuration) linked to the contributions of the modulation-independent and modulation-dependent nonlinearities, respectively.
The expression in \eqref{nli} includes traditional inter- and  intra-channel effects
as well as additional intra-channel terms for signal with
dense WDM, as discussed in \cite{Carena:14}. The four nonlinear terms $\eta_1$,  $\eta_2$, $\eta_3$ and  $\eta_4$ can be calculated via Monte Carlo integration from the ready-to-use MATLAB code presented in \cite{Dar:14}.

The coefficient $\mu_4$ and $\mu_6$ in \eqref{nli} are the fourth and sixth standardized moments of the constellation used. In general, the $p$-th standardized moment
$\mu_p$ of the channel input $\bX$ is defined as
\begin{align}\label{eq:standardized_moments}
\mu_p=\frac{\mathbb{E}\left[\|\bX-\mathbb{E}[\bX]\|^p\right]}{\left(\mathbb{E}\left[\|\bX-\mathbb{E}[\bX]\|^2\right]\right)^{\frac{p}{2}}}.
\end{align}
From \eqref{nli} we can observe that NLIN is dependent on the modulation format.  
\cite{SerenaECOC2013} shows that $M$PSK with constant-modulus can significantly reduce the NLI noise, which yield $\mu_4=1$ and $\mu_6=1$.
However, performance of $M$PSK in terms of AIRs for the AWGN is poor because the amplitude of the symbols does not carry information. Half of the degrees of freedom in 2D space are therefore lost.
On the other hand, either PS or GS with enough constellation points offer shaping gains for the AWGN channel of up to $1.53$~dB. This is achieved by using a continuous Gaussian distribution, which in turn yields a larger modulation-dependent NLI ($\mu_4=2$ and $\mu_6=6$).
 Gaussian distributions are therefore optimal for the AWGN channel but result in a larger penalty in the nonlinear optical channel. PS with different input distributions have been investigated to trade-off between linear and nonlinear shaping gain \cite{PanJLT2016,YankovnJLT2016,RennerJLT2017,RasmusArxiv2018,SillekensOFC2018}. 
{Meanwhile, the obtained nonlinear shaping gains of two-dimensional GS are limited because there are limited degrees of freedom for optimization in 2D space \cite{SillekensECOC2018}}.

To achieve larger shaping gains, more than two dimensions need to be used. The design should also aim at ``shaping-out" the modulation-dependent NLIN term, as explained in \cite{DarISIT2014}. To correctly shape-out the NLIN, a precise multidimensional model is required. However, current NLIN models available in the literature can only predict the modulation-dependent NLIN term of a polarization-multiplexed system with independent modulation formats over two polarization \cite{DarJLT2015}. One exception to this is the general expression presented in \cite[Eq. (3)]{DarJLT2015}, which gives an intuition that constant-modulus across two polarization can in fact reduce the induced NLIN due to self-phase modulation (SPM) and cross-phase modulation (XPM). At the same time, 4D constant-modulus formats have also been reported to show a significant reduction in SPM and XPM by separating nonlinear components. This was done via numerical simulations in \cite[Sec.~IV]{Kojima2017JLT}. In this paper, the constant-modulus constraint for a 4D modulation will be taken into account.

\subsection{Design Strategy and Optimization}\label{sec:Format}

The theoretical analysis of the previous sections suggests that the MD modulation format design should maximize linear shaping gain (i.e., approach a Gaussian-like modulation) and at the same time minimize the shaping penalty due to the increased NLI (i.e., approach a constant-modulus modulation).
Based on these observations, our optimized 4D modulation formats are designed in two rules:
\begin{enumerate}[leftmargin=5ex,itemsep=0.5ex,label={[R\arabic*]}]
\item \textbf{Maximize Linear Performance}: Numerically optimize 4D modulation formats so that the GMI is maximized. This is achieved by jointly optimizing the 4D coordinates and the binary labeling. 
\item \textbf{Maximize Nonlinear Performance}: Add a constraint of constant  modulus in 4D to reduce the nonlinear penalty.
\end{enumerate}

In this paper, we consider constellations with SEs of $6$~bit/4D-sym, which are comparable with PM-8QAM. The modulation format reported below includes a constant-modulus constraint, which  cause a small loss with respect to the best achievable linear performance. This constraint, however, results in considerably better nonlinear performance. We believe this methodology can be also extended to designing modulation formats with SEs between $7$ and $10$~bit/4D-sym by using quasi-constant modulus to trade-off shaping gain and nonlinear penalty.

The GMI-based optimization problem we solve in this paper corresponds to find a constant-modulus constellation $\mathbb{S}$ and labeling $\mathbb{B}$ for a given noise variance $\sigma_z^2$  under an energy constraint, i.e., 
\begin{align}\label{eq:OP_GMI}
\{\mathbb{S}^*,\mathbb{B}^*\} & =\argmax_{\mathbb{S},\mathbb{B}} \{G(\sigma_z^2,\mathbb{S},\mathbb{B})\},\\ \label{eq:OP_GMI2}
\text{subject to~:~}
\|\bs_i\|^2& =E_s\quad \forall i\in\{1,2,\ldots,M\}\\ \nonumber
E_s & \leq \sigma_x^2
\end{align}
where $\mathbb{S}^*$ and $\mathbb{B}^*$ indicate the optimal constellation and labeling, resp. {Note that the optimization problem is a single objective function (GMI maximization) with multiple parameters  and constraints.}

\begin{figure}[tbp]
\vspace{-1em}
    \centering
   \includegraphics[width=\columnwidth]{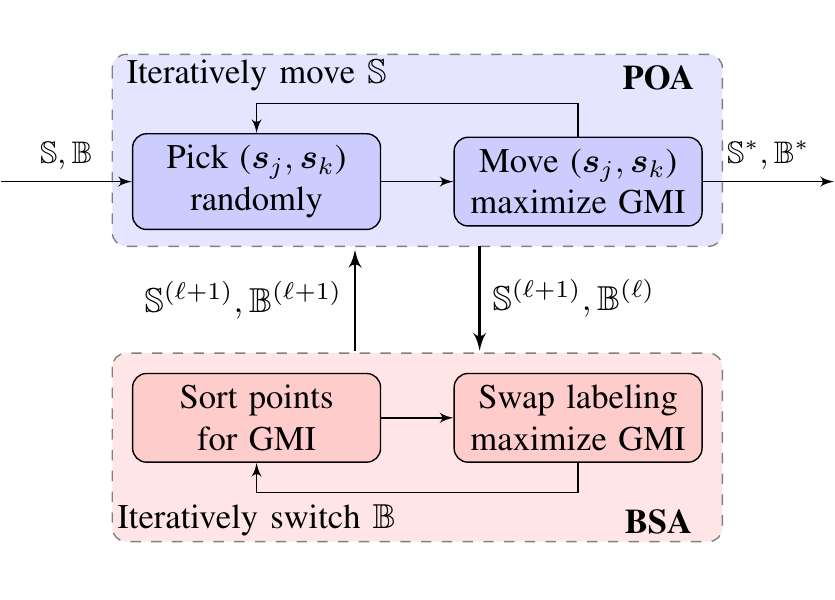}
   \vspace{-1em}
    \caption{Schematic  diagram  of  the  optimization  algorithm: $(\bs_{j},\bs_{k})$ is the selected points as a pair in each POA iteration, $\ell$ is the index of outer iteration number between POA and BSA.}
    \label{fig:optimization_model}
\end{figure}

{In this paper, we solve \eqref{eq:OP_GMI}--\eqref{eq:OP_GMI2} using the pairwise optimization
algorithm (POA) \cite{Moore2009,ZhangOFC2017,ZhangECOC2017} to optimize the coordinates $\mathbb{S}$ and the binary switch algorithm (BSA) \cite{Schreckenbach2003} to optimize the labeling $\mathbb{B}$.
The schematic diagram of the optimization algorithm is shown in Fig. \ref{fig:optimization_model}. 
The optimization first implement POA (blue area) with  randomly chosen coordinates and  labeling. After a given number of POA iterations, the BSA (red area)  with the GMI approximation in \cite{AlvaradoCommL2014} is applied to avoid the optimization to converge to a local optimum. 
The entire optimization process is also implemented in an iterative fashion between POA and BSA with outer iteration index $\ell$, and is repeated until the algorithm has converged.
\footnote{{Note that in this paper, we are interested in using the correct objective function to find an optimum (or close to optimum) solution for a given system with SD-FEC overhead of $20\%-25\%$. Other optimization methods can also be used to find the optimal solution. This includes for example genetic algorithms \cite{El-RahmanECOC2017} and machine learning approaches \cite{ShenLIECOC2018,RasmusArxiv2018}.}}
The POA implements as following steps: 
1) Configure the initial constellation.\footnote{{Initial constellations does affect the time until convergence. In this paper, we use PM-8QAM as initial constellation.  A detailed impact of initial constellations, which we leave to future studies, is of great practical interest.}} 2) Select a pair of points $(\bs_{j},\bs_{k})$ in coordinates set $\mathbb{S}$. 3) Find the position of the pair of points $(\bs_{j},\bs_{k})$ maximizing GMI. 4) Go back to Step 2 and repeat until the maximum number of iteration is achieved. 
BSA iteratively improves the labeling by switching the mapping of points. 
In each iteration of the searching process, two labeling indices of symbols with  the highest cost in the list  are switched with each other after an ordered list of points are generated.
If the switching generates a lower cost than the cost before the labeling switch, then the switch is accepted, otherwise the switch is ignored.
Note that there are many local optima  for large constellation and/or for constellation with high dimensionality, which has been reported by \cite{ZhangOFC2017}.
To overcome this problem, we select the pairs in the POA randomly and combine it iteratively with the BSA.
}

Our optimization was done without any symmetry considerations. However, in order to speed up the optimization, symmetry constraints can be used.  In particular, we forced all 16 orthants to be identical.\footnote{An orthant in $N$-dimensions is the intersection of $N$ mutually orthogonal half-spaces. By independent selections of half-space signs, there are $2^N$ orthants in $N$-dimensional space.} Furthermore, $4$ out of the $6$ bits were assigned to determine the orthants, and the labeling optimization was done only over the remaining two bits. 
{By using the symmetry constraints, the optimization is simplified. Only four 4D coordinates and the binary labeling need to be found, which significantly reduces the convergence time.}
 Interestingly, both the unconstrained optimization and the one with the symmetries gave the same result {at the optimized SNR of 8~dB.}
 {The intuition behind the symmetry property is that the GMI-based  constellation optimization try to balance the 
trade-off between the number of the nearest neighbors and the distance between symbols with different Hamming
distance. In order to reduce the mapping penalty with respect to  Gray mapping, a symmetric structure can attempt to make all the neighbouring symbols with one bit difference to have equal distance.
However, for MI-based constellation optimization, the optimum solutions do not necessarily have the symmetric property \cite{BinECOC2018}.}
{In addition, the symmetry property reduces the complexity of soft-demapping.}

\subsection{Proposed Format: 4D-64PRS}

The proposed 6 bit/4D-sym modulation format with $M=2^m=64$ points is labeled by $m=6$ bits. We call this new modulation format with its binary labeling four-dimensional polarization-ring-switching 4D-64PRS. The coordinates of this family of modulation formats and the corresponding binary labeling are given in Table~\ref{tab:4D_64_XL}.  As shown in Table~\ref{tab:4D_64_XL}, the coordinates of 4D-64PRS are such
\begin{align}\label{eq:alphabet}\nonumber
\bs_i \in \{&[\pm \nu_1, \pm \nu_3, \pm \nu_2, \pm \nu_2], [\pm \nu_3, \pm \nu_1, \pm \nu_2, \pm \nu_2],\\
&[\pm \nu_2, \pm \nu_2, \pm \nu_1, \pm \nu_3], [\pm \nu_2, \pm \nu_2, \pm \nu_3, \pm \nu_1]\},
\end{align}
and therefore, the proposed format is highly symmetric.

\begin{table}[tbp]
\caption{Coordinates and binary labeling of the 4D-64PRS format.}
\label{tab:4D_64_XL}
\centering
\scalebox{1.0}{
\centering {\footnotesize
\begin{tabular}{@{}|@{\hskip 0.5ex}c@{\hskip 0.5ex}|@{\hskip 0.5ex}c@{\hskip 0.5ex}|@{\hskip 0.5ex}c@{\hskip 0.5ex}|@{\hskip 0.5ex}c@{\hskip 0.5ex}|@{}}
\hline

\hline
 Coordinates & Labeling & Coordinates & Labeling\\ 
\hline 

\hline
$(+\nu_1,+\nu_3,+\nu_2,+\nu_2)$ & 000000&  
$(+\nu_1,+\nu_3,-\nu_2,+\nu_2)$ & 000010\\ \hline  
$(+\nu_1,+\nu_3,-\nu_2,-\nu_2)$ & 000110&  
$(+\nu_1,+\nu_3,+\nu_2,-\nu_2)$ & 000100\\ \hline  
$(-\nu_1,+\nu_3,+\nu_2,+\nu_2)$ & 010000&  
$(-\nu_1,+\nu_3,-\nu_2,+\nu_2)$ & 010010\\ \hline  
$(-\nu_1,+\nu_3,-\nu_2,-\nu_2)$ & 010110&  
$(-\nu_1,+\nu_3,+\nu_2,-\nu_2)$ & 010100\\ \hline  
$(-\nu_3,+\nu_1,+\nu_2,+\nu_2)$ & 011001&  
$(-\nu_3,+\nu_1,-\nu_2,+\nu_2)$ & 011011\\ \hline  
$(-\nu_3,+\nu_1,-\nu_2,-\nu_2)$ & 011111&  
$(-\nu_3,+\nu_1,+\nu_2,-\nu_2)$ & 011101\\ \hline  
$(-\nu_3,-\nu_1,+\nu_2,+\nu_2)$ & 111001&  
$(-\nu_3,-\nu_1,-\nu_2,+\nu_2)$ & 111011\\ \hline  
$(-\nu_3,-\nu_1,-\nu_2,-\nu_2)$ & 111111&  
$(-\nu_3,-\nu_1,+\nu_2,-\nu_2)$ & 111101\\ \hline  
$(-\nu_1,-\nu_3,+\nu_2,+\nu_2)$ & 110000&  
$(-\nu_1,-\nu_3,-\nu_2,+\nu_2)$ & 110010\\ \hline  
$(-\nu_1,-\nu_3,-\nu_2,-\nu_2)$ & 110110&  
$(-\nu_1,-\nu_3,+\nu_2,-\nu_2)$ & 110100\\ \hline  
$(+\nu_1,-\nu_3,+\nu_2,+\nu_2)$ & 100000&  
$(+\nu_1,-\nu_3,-\nu_2,+\nu_2)$ & 100010\\ \hline  
$(+\nu_1,-\nu_3,-\nu_2,-\nu_2)$ & 100110&  
$(+\nu_1,-\nu_3,+\nu_2,-\nu_2)$ & 100100\\ \hline  
$(+\nu_3,-\nu_1,+\nu_2,+\nu_2)$ & 101001&  
$(+\nu_3,-\nu_1,-\nu_2,+\nu_2)$ & 101011\\ \hline  
$(+\nu_3,-\nu_1,-\nu_2,-\nu_2)$ & 101111&  
$(+\nu_3,-\nu_1,+\nu_2,-\nu_2)$ & 101101\\ \hline  
$(+\nu_3,+\nu_1,+\nu_2,+\nu_2)$ & 001001&  
$(+\nu_3,+\nu_1,-\nu_2,+\nu_2)$ & 001011\\ \hline  
$(+\nu_3,+\nu_1,-\nu_2,-\nu_2)$ & 001111&  
$(+\nu_3,+\nu_1,+\nu_2,-\nu_2)$ & 001101\\ \hline  
$(+\nu_2,+\nu_2,+\nu_1,+\nu_3)$ & 001000&  
$(+\nu_2,+\nu_2,-\nu_1,+\nu_3)$ & 001010\\ \hline  
$(+\nu_2,+\nu_2,-\nu_3,+\nu_1)$ & 000011&  
$(+\nu_2,+\nu_2,-\nu_3,-\nu_1)$ & 000111\\ \hline  
$(+\nu_2,+\nu_2,-\nu_1,-\nu_3)$ & 001110&  
$(+\nu_2,+\nu_2,+\nu_1,-\nu_3)$ & 001100\\ \hline  
$(+\nu_2,+\nu_2,+\nu_3,-\nu_1)$ & 000101&  
$(+\nu_2,+\nu_2,+\nu_3,+\nu_1)$ & 000001\\ \hline  
$(-\nu_2,+\nu_2,+\nu_1,+\nu_3)$ & 011000&  
$(-\nu_2,+\nu_2,-\nu_1,+\nu_3)$ & 011010\\ \hline  
$(-\nu_2,+\nu_2,-\nu_3,+\nu_1)$ & 010011&  
$(-\nu_2,+\nu_2,-\nu_3,-\nu_1)$ & 010111\\ \hline  
$(-\nu_2,+\nu_2,-\nu_1,-\nu_3)$ & 011110&  
$(-\nu_2,+\nu_2,+\nu_1,-\nu_3)$ & 011100\\ \hline  
$(-\nu_2,+\nu_2,+\nu_3,-\nu_1)$ & 010101&  
$(-\nu_2,+\nu_2,+\nu_3,+\nu_1)$ & 010001\\ \hline  
$(-\nu_2,-\nu_2,+\nu_1,+\nu_3)$ & 111000&  
$(-\nu_2,-\nu_2,-\nu_1,+\nu_3)$ & 111010\\ \hline  
$(-\nu_2,-\nu_2,-\nu_3,+\nu_1)$ & 110011&  
$(-\nu_2,-\nu_2,-\nu_3,-\nu_1)$ & 110111\\ \hline  
$(-\nu_2,-\nu_2,-\nu_1,-\nu_3)$ & 111110&  
$(-\nu_2,-\nu_2,+\nu_1,-\nu_3)$ & 111100\\ \hline  
$(-\nu_2,-\nu_2,+\nu_3,-\nu_1)$ & 110101&  
$(-\nu_2,-\nu_2,+\nu_3,+\nu_1)$ & 110001\\ \hline  
$(+\nu_2,-\nu_2,+\nu_1,+\nu_3)$ & 101000&  
$(+\nu_2,-\nu_2,-\nu_1,+\nu_3)$ & 101010\\ \hline  
$(+\nu_2,-\nu_2,-\nu_3,+\nu_1)$ & 100011&  
$(+\nu_2,-\nu_2,-\nu_3,-\nu_1)$ & 100111\\ \hline  
$(+\nu_2,-\nu_2,-\nu_1,-\nu_3)$ & 101110&  
$(+\nu_2,-\nu_2,+\nu_1,-\nu_3)$ & 101100\\ \hline  
$(+\nu_2,-\nu_2,+\nu_3,-\nu_1)$ & 100101&  
$(+\nu_2,-\nu_2,+\nu_3,+\nu_1)$ & 100001\\ \hline  

\hline 

\hline
\end{tabular}} 
}
\end{table} 

\begin{figure}[tbp]
\centering
\includegraphics[width=0.48\columnwidth]{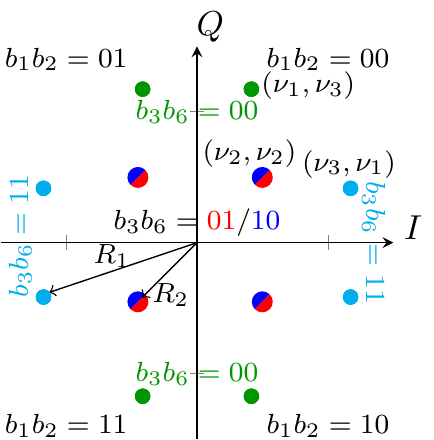}
\includegraphics[width=0.48\columnwidth]{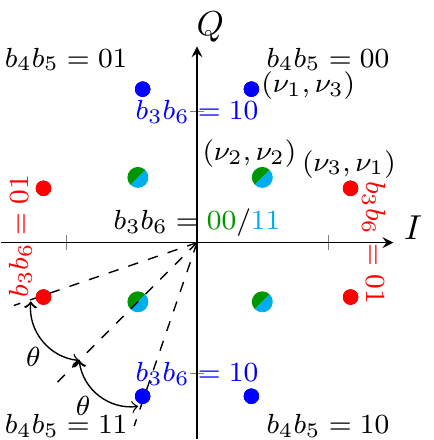}
\caption{2D-projections of the designed 4D-64PRS modulation and associated bits/polarization bits mapping. The rings are given by $R_1^2={\nu_1^2+\nu_3^2}$ and $R_2^2=2\nu_2^2$.}
\label{fig:4D_64_modulation_label}
\end{figure}

The 4D-64PRS format has 64 non-overlapping points in 4D space. For better visualization, these points can be projected on the first two and last two dimensions (i.e., on the two polarizations). This projection results in 12 distinct points in each 2D space, as shown in Fig.~\ref{fig:4D_64_modulation_label}. In order to clearly show the 4D-64PRS modulation projected on 2D, and also to emphasize the inter-polarization dependency, we use in Fig.~\ref{fig:4D_64_modulation_label} the following color coding strategy: 2D projected symbols in the first and second polarization are valid 4D symbols only if they share the same color. Consider for example the $4$ blue points in first 2D and the $4$ blue points in second 2D, each of the $4\times4=16$ combinations correspond to a 4D symbol. However, a blue point in the first 2D cannot be transmitted in combination with a green/red/light blue point in the second 2D. The same rules applies to light blue, green, and red points, giving a total of $16 \times 4=64$~4D points.

The color coding scheme used in Fig.~\ref{fig:4D_64_modulation_label} also shows the constant-modulus constraint we imposed on the format, where $R_1$ and $R_2$ are the ring radii. The 4D-64PRS format can be seen as the transmission of alternating rings in both polarizations. In other words, when an inner ring ($R_2$) is transmitted in the first 2D (first polarization), the outer ring ($R_1$) is used in the second 2D (second polarization). This case corresponds to blue or red points. The opposite alternation (outer ring, inner ring) is used for light blue or green points. This ``polarization-ring-switching'' explains the name we gave to this format. As mentioned above, a feature of the 4D-64PRS is its symmetry. Fig.~\ref{fig:4D_64_modulation_label} shows that the projections of 4D-64PRS on both polarizations are identical. Furthermore, the points in each quadrant are simply rotated versions of the points
{$(\nu_1,\nu_3)$, $(\nu_2,\nu_2)$, and $(\nu_3,\nu_1)$}
in the first quadrant.

\begin{figure}[tbp]
\centering
\includegraphics[width=\columnwidth]{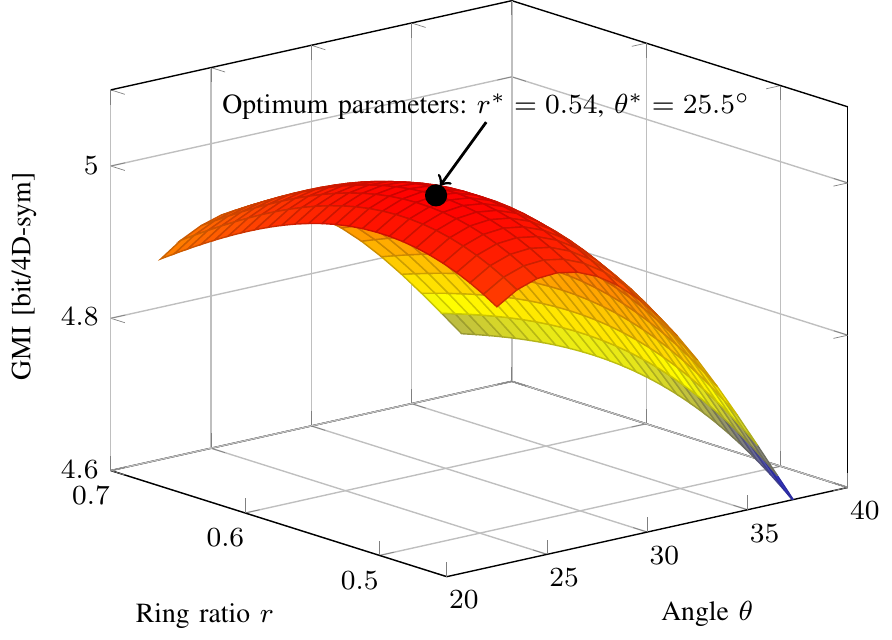}
\vspace{-2em}
\caption{GMI as a function of ring ratio $r$ and angle $\theta$, SNR=$8$~dB in AWGN for 4D-64PRS. The optimum parameters are highlighted.}
\label{fig:paramtersOpt}
\end{figure}

The family of 4D-64PRS formats we propose assumes that the symmetries in terms of the orthants and angle between the points are maintained. The format can then be optimized in terms of two parameters: the ring ratio $r=R_2/R_1$ and angle $\theta$, shown in Fig.~\ref{fig:4D_64_modulation_label}. The GMI as a function of $r$ and $\theta$ at SNR=$8$~dB are shown in Fig. \ref{fig:paramtersOpt}. The GMI for this SNR is clearly concave on $r$ and $\theta$, and thus, its optimization is trivial. The optimum values for SNR=$8$~dB are $r^*=0.54$ and $\theta^*=25.5\degree$, which are highlighted in Fig. \ref{fig:paramtersOpt}. The optimization for SNR=$8$~dB can be understood as an optimization for an effective SNR in \eqref{eq:eff} of $8$~dB which ignores the NLI. The resulting AIR is approximately $5$~bit/4D-sym, i.e., we are targeting a FEC rate of $R=5/6$. The coordinates of the constellation in Table~\ref{tab:4D_64_XL} are $\nu_1=0.87, \nu_2=1, \nu_3=2.47$ (rounded to two decimal points).

\begin{figure}[tbp]
\centering
\includegraphics[width=\columnwidth]{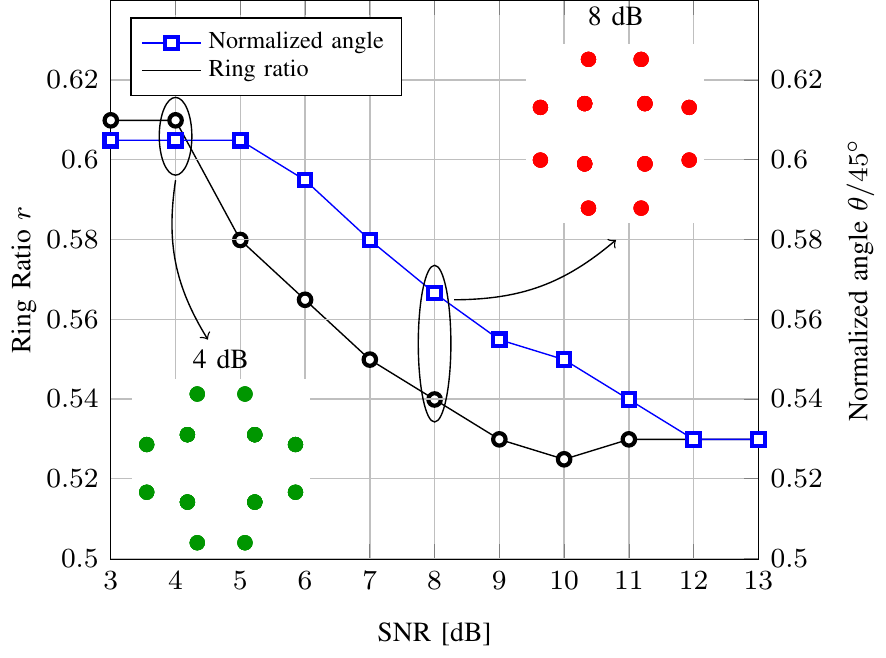}
\vspace{-2em}
\caption{Optimized ring ratio $r$ and  normalized angle $\theta/45\degree$ vs SNR in AWGN for 4D-64PRS modulation. {Insets: the optimized 4D modulation formats in 2D-projection for SNR= $\{ 4, 8 \}$~dB.}}
\label{fig:paramtersOpt_SNR}
\end{figure}

We have also optimized the 4D-64PRS format for a range of SNR by optimizing $r$ and $\theta$, as shown in Fig. \ref{fig:paramtersOpt_SNR}.  
{The  optimized  4D  modulation  formats  in 2D-projection for SNR=$\{4,8\}$~dB are also shown as insets.}
This figure shows that at different SNR regions, different $r$ and $\theta$ should be chosen to maximize the GMI. In particular, we observe small variations of the optimal angle (between $27.2\degree$ and $23.4\degree$), with a general tendency to decrease as the SNR increases. The ring ratio also shows small variations, decreasing from $0.61$ to $0.53$ as the SNR increases.

Regarding the binary labeling of 4D-64PRS, we show in Fig.~\ref{fig:4D_64_modulation_label} that the quadrants in each polarization are defined by 4 out of 6 bits, namely, $[b_1,b_2]$ and $[b_4,b_5]$ for the first and second polarization, respectively. These $4$ bits can be seen as the ones determining which of the 16 orthants is to be chosen. The remaining 2 bits [$b_3,b_6$] determine the symbol from the 4 possible constellation points in the same orthant.
In other words, [$b_3,b_6$] determine the color of the transmitted points in Fig. \ref{fig:4D_64_modulation_label} while [$b_1,b_2,b_4,b_5$] determine the coordinate of the point in the same color.

\begin{figure*}[!b]
\centering
 \includegraphics[width=0.95\textwidth]{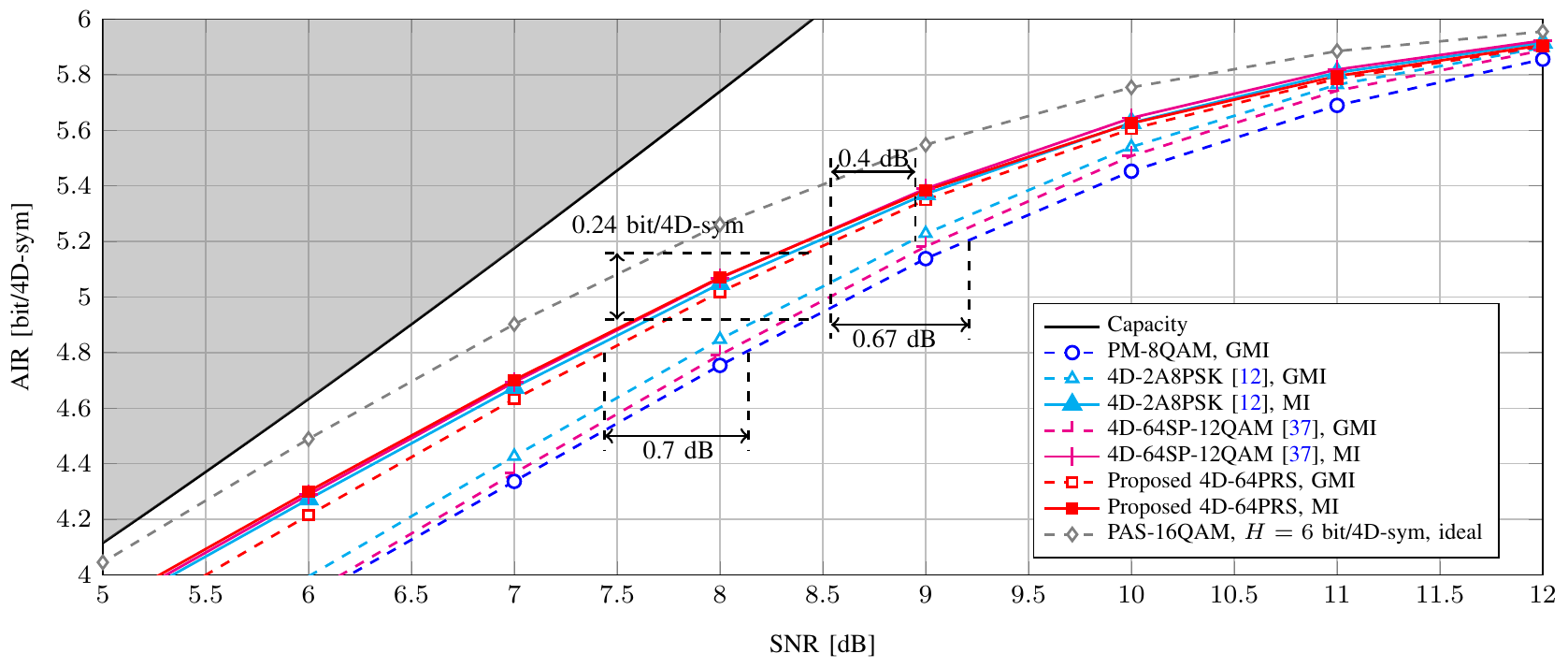}
  \caption{MI and GMI as a function of SNR for various modulation formats with 6~bit/4D-sym. The  PM-8QAM, 4D-64SP-12QAM and 4D-2A8PSK are  shown as a baseline.}
\label{fig:GMI_4D_64}
\end{figure*}

We conclude this section by discussing the differences between 4D-64PRS and other existing 4D modulation formats. The key design difference with respect to set-partitioned 16QAM \cite{RenaudierECOC2012,NakamuraECOC2015} and 4D-2A8PSK \cite{Kojima2017JLT}, is that we jointly optimize the labeling and coordinates of the constellation points to achieve higher GMI (see next Sec. for details). When projected on 2D,  4D-64PRS has a similar shape to 4D-64SP-12QAM \cite{NakamuraECOC2015}. However, 4D-64PRS has a different 4D geometrical structure and different labeling, which cannot be obtained by set-partitioning from PM-16QAM. 
As we will see in the next section, 4D-64PRS and its optimized labeling provides significant linear shaping gains with respect to 8QAM, 4D-64SP-12QAM and 4D-2A8PSK. Furthermore, the constant-modulus property of 4D-64PRS  yields excellent nonlinear transmission performance.

\section{Numerical Results}\label{Sec:NumericalResults}

In the following, we numerically evaluate the AIRs for both the AWGN channel and a multi-span fiber system. Different modulation formats with $6$~bit/4D-sym are considered. The analysis focuses on linear shaping gains and the effect of  shaping on the fiber nonlinearities and effective SNR. 

\subsection{Results for AWGN Channel}\label{Sec:NumericalResultsAWGN}

Fig.~\ref{fig:GMI_4D_64} shows the linear performance in terms of AIRs for three 4D modulation formats with $6$~bit/4D-sym: 4D-2A8PSK, 4D-64SP-12QAM and the proposed 4D-64PRS.\footnote{{Note that to achieve different rate points on the AIR curves requires variable-rate FEC.}} The 4D-64PRS format we use was optimized for SNR=$8$~dB . An optimized ring ratio of $0.65$ was used for 4D-2A8PSK. The results in Fig.~\ref{fig:GMI_4D_64} show that the three 4D modulation formats have a similar MI performance  (solid lines). This figure also show that the GMI of 4D-28PSK and 4D-64SP-12QAM is considerably lower than their corresponding MI, as previously shown for other 4D formats in \cite{Alvarado2015_JLT}. 
{On the other hand, the gap between the GMI and MI of 4D-64PRS is relatively small. 
This indicates that the labeling penalty  of the proposed 4D-64PRS is negligible.
}

Considering the GMI of PM-8QAM as a baseline, 4D-64PRS can provide  gains of $0.7$~dB at an AIR between $4.8$ and $5.2$~bit/4D-sym. AIR gains of approximately $0.24$~bit/4D-sym are observed for this AIR range. The 4D-64PRS format also perform better than 4D-2A8PSK \cite{Kojima2017JLT} and 4D-12QAM \cite{NakamuraECOC2015} (in terms of GMI) by at least 0.4 dB.

{Probabilistic amplitude shaped (PAS)-16QAM with entropy $H=6~$bit/4D-sym was generated using Maxwell-Boltzmann distributions. This GMI is an upper bound for the GMI of PAS with an ideal infinite block length distribution matcher (DM). 
Even probabilistic shaping allows to operate closely to the AWGN capacity,  it will lead to significant rate loss  for distribution matcher with short block length.} 

\begin{figure}[!tb]
\centering
\vspace{-0.5em} 
 \includegraphics[width=\columnwidth]{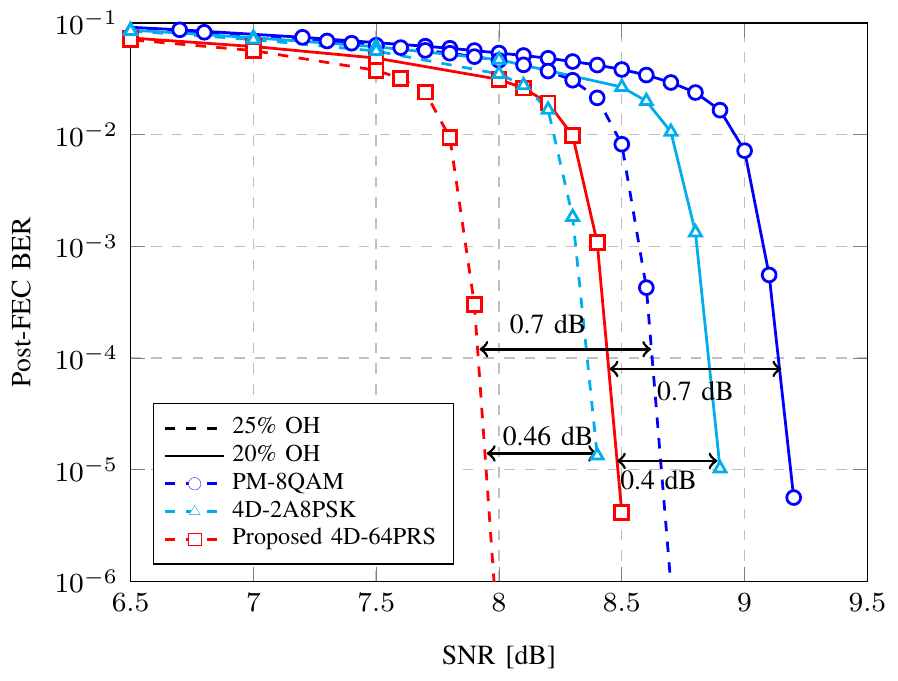}
\caption{Post-FEC BER performance of formats with $6$~bit/4D-sym and {LDPC with} 20\% and 25\% FEC overhead.}
\label{fig:GMIpostBER}
\end{figure}

In order to verify the accuracy of GMI prediction, we simulated the  low-density parity-check (LDPC) codes used in DVB-S2 standard. FEC overheads of $20\%$ and $25\%$ with 64800 bits per frame for PM-8QAM, 2A8PSK and 4D-64PRS were used. The results are shown in Fig. \ref{fig:GMIpostBER}. We can observe that the gains for 4D-64PRS in terms of post-FEC BER coincide well to the GMI prediction in Fig.~\ref{fig:GMI_4D_64}.

\subsection{Results for Nonlinear Fiber Optical Channel}\label{Sec:NumericalResultsNLIFiber}

The 4D modulation formats 4D-2A8PSK, 4D-64SP-12QAM and 4D-64PRS were implemented in two polarizations and simulated  over the optical fiber channel.
The optical fiber channel comprising multiple standard single-mode
fiber spans with $\alpha=0.21$~dB/km, $D=16.9$~ps/nm/km, $\gamma=1.3175$ $($W$\cdot$km$)^{-1}$. 
A dual-polarization multi-span WDM system with 11 co-propagating channels
was transmitted at a symbol rate of $45$~GBaud, a WDM spacing of $50$~GHz  and a root-raised-cosine (RRC) filter roll-off factor of $0.1$. 
Each WDM channel carries independent data with  $N_\text{sym}=2^{16}$ and are transmitted at the same power.
 Each span of length $80$~km was followed by an erbium-doped fiber amplifier (EDFA) with a noise figure of $5$~dB. Note that polarization mode dispersion (PMD) was not considered.
At the receiver, an ideal receiver is used for detection: chromatic dispersion is digitally compensated, then the signal is matched filtered and downsampled. Potential constant phase rotations are ideally compensated. 


\begin{figure}[!tb]
\centering
   \includegraphics[width=\columnwidth]{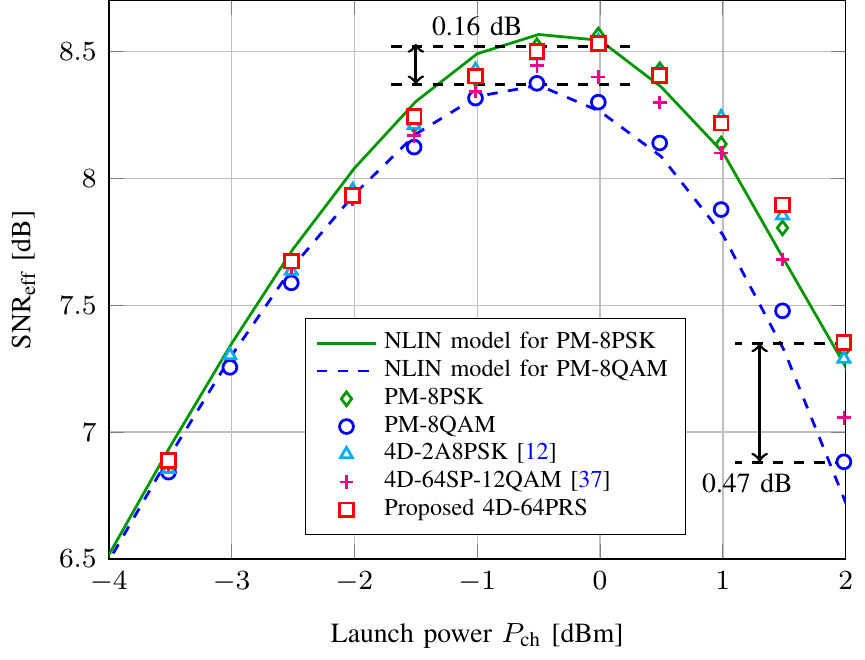}
  \caption{Effective SNRs $\text{SNR}_{\text{eff}}$ as function of the transmitted power for different modulation formats with 6~bit/4D-sym over a 8000 km link. 
}
\label{fig:SNR_P_4D_64}
\vspace{-0.3em} 
\end{figure}

We compare the effective SNR (after fiber propagation and receiver DSP) for an 8000 km link. The results are shown in Fig. \ref{fig:SNR_P_4D_64}.  Results for the numerical simulations  (markers) and the estimated SNR by NLIN model \cite{Dar:14} for PM-8QAM (dashed lines)  and PM-8PSK (solid lines) are shown. A good agreement between the simulation and NLIN model in \cite{Dar:14} is observed. The simulations results in Fig. \ref{fig:SNR_P_4D_64} show that the two 4D modulation formats under consideration (4D-2A8PSK and 4D-64PRS) provide an effective SNR $0.16$~dB higher than PM-8QAM.
Since the design of 4D-64SP-12QAM does not consider the modulation-dependent NLI, its effective SNR is worse than 4D-2A8PSK and the proposed 4D-64PRS, but it is a little bit higher than PM-8QAM.
In addition, Fig. \ref{fig:SNR_P_4D_64}  also shows the effective SNR of PM-8PSK with constant-modulus as a baseline, which has the smallest modulation-dependent NLI. We observe that the three constant-modulus modulation formats have similar effective SNR, and PM-8QAM and 4D-64SP-12QAM with power fluctuation between time slots have an effective SNR penalty due to fiber nonlinearities. The penalty becomes 0.47~dB when the launch power increase. 
{At the optimum launch power, the proposed modulation format  partially reduces fiber nonlinear interference (the modulation-dependent NLI) and therefore increases the effective SNR with respect to PM-8QAM}. The total shaping gain is linear shaping gain plus NLI shaping gain. Below we will show  how this translates into a reach increase.

\begin{figure}[!tb]
\centering
   \includegraphics[width=\columnwidth]{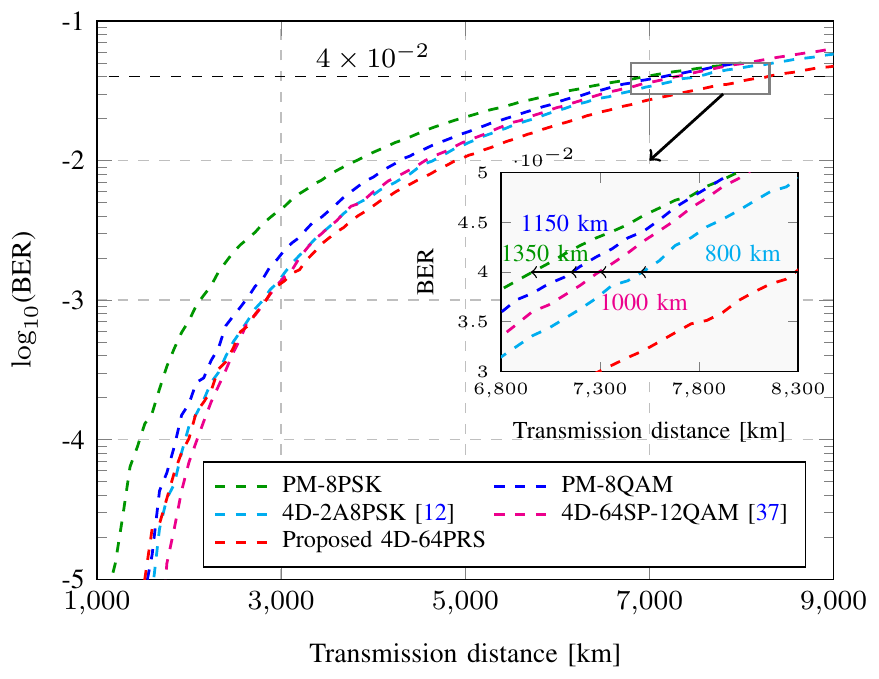}
  \caption{BER as function of the transmission distance for different 4D modulation formats with 6~bit/4D-sym using the optimal launch power at each distance.}
\label{fig:BER_D_4D_64}
\end{figure}

Fig. \ref{fig:BER_D_4D_64} shows pre-FEC BER 
as a function of the transmission distance for different modulation formats  using the optimal launch power at each distance. At BER= $4\times 10^{-2}$, which is the typical
BER threshold of the state-of-the-art SD-FEC having a code rate of 0.8, 
the achievable transmission distance is 8300 km for 4D-64PRS.
4D-64PRS yields a   1350 km  and 1150 km reach increases  with respect to  PM-8PSK and PM-8QAM, resp. 4D-64PRS also provides 1000 km and 800 km gain compared  to 4D-64SP-12QAM and  4D-2A8PSK, resp.
For pre-FEC BER below $1\times 10^{-3}$, 4D-64SP-12QAM provides the best performance. 
It is not only because 4D-64SP-12QAM has largest minimum Euclidean, but also {because} the 4D-64PRS is not designed for such higher SNR at  short distance.

\begin{figure}[!tb]
\centering
   \includegraphics[width=\columnwidth]{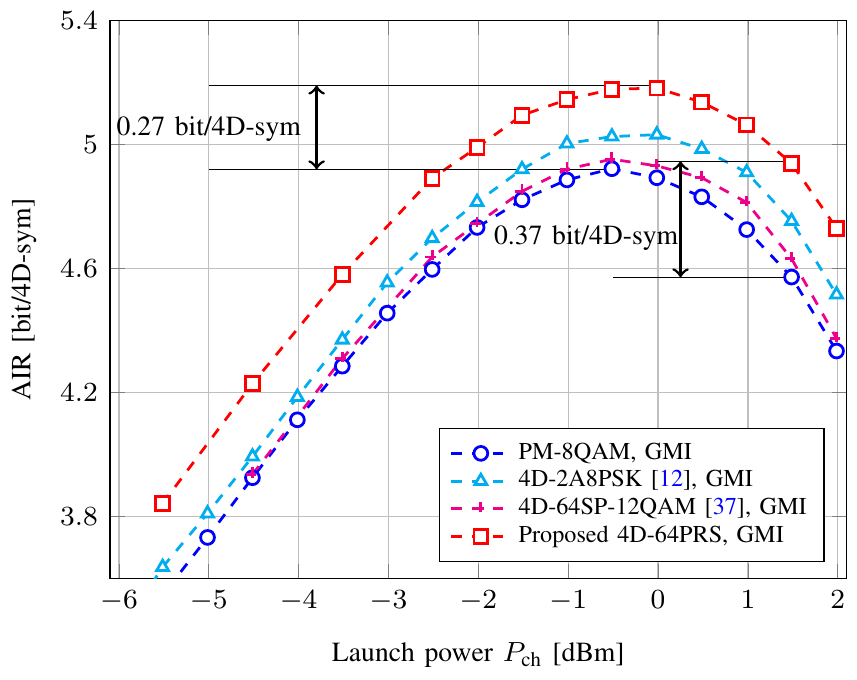}
  \caption{AIRs as function of the transmitted
power for different modulation formats with 6~bit/4D-sym over a 8000 km link.
}
\label{fig:GMI_P_4D_64}
\end{figure}

Fig. \ref{fig:GMI_P_4D_64} shows AIRs as a function of the transmitted power for different modulation formats with 6~bit/4D-sym over a 8000 km link.
Around the optimum launch power regime  where linear  and nonlinear propagation effects are comparable,
4D-64PRS modulation gives 0.27 bit/4D-sym GMI gain with respect to PM-8QAM. 
For higher launch power regime where nonlinearity is dominant, performance improvements become more significant, and this GMI gain is around 0.37 bit/4D-sym. 
It is important to note that the GMI gain in the nonlinear channel (at the optimum launch power, 0.27~bit/4D-sym) is higher than the linear channel in Fig. \ref{fig:GMI_4D_64} (0.24~bit/4D-sym). This additional gain (or equivalently, the almost absence of NLIN penalty) validates the efficiency of the NLI tolerant design of the proposed 4D-64PRS format. 
In Fig. \ref{fig:GMI_P_4D_64}, we also plot the AIRs of other 4D modulation formats (4D-2A8PSK and 4D-64SP-12QAM). We can observe that 4D-64PRS outperforms both of them in terms of GMI due to its excellent linear performance and also the nonlinear-tolerant property.

\begin{figure*}[!tb]
\centering
   \includegraphics[width=0.95\textwidth]{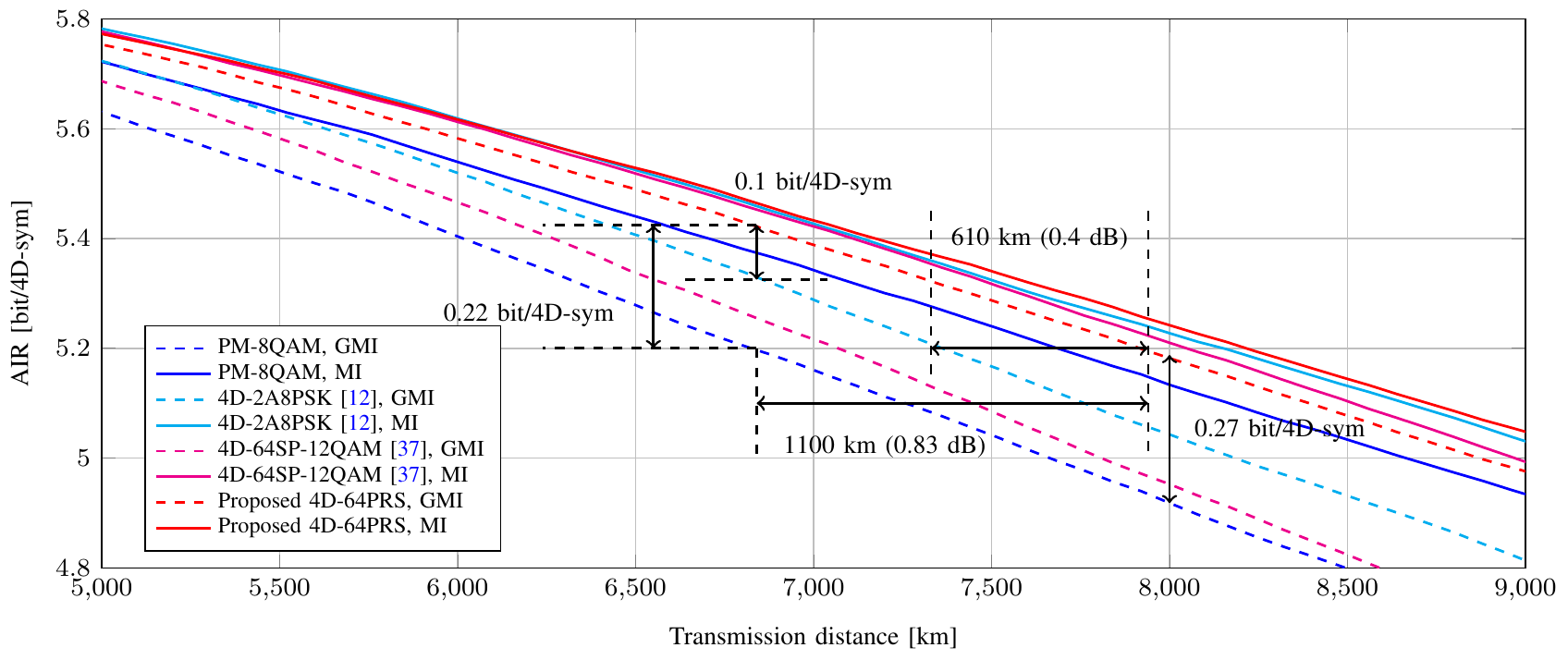}
  \caption{AIRs as function of the transmission distance for different 4D modulation formats with 6~bit/4D-sym.}
\label{fig:GMI_D_4D_64}
\end{figure*}

To conclude this section, Fig. \ref{fig:GMI_D_4D_64} shows AIRs as a function of the transmission distance using the optimal launch power at each distance. 4D-64PRS yields a 1100 km increase in reach relative to PM-8QAM at GMI of 5.2 bit/4D-sym, and more than 610 km relative to 4D-2A8PSK. At a transmission distance of 6840 km, 4D-64PRS also provides 0.22 bit/4D-sym and 0.1 bit/4D-sym gain compared to PM-8QAM and 4D-2A8PSK, resp.

The AIR gains from shaping translate into sensitivity improvements in nonlinear optical channel, which is linear gain plus the nonlinear gain. {In particular, the reach increases in Fig. \ref{fig:GMI_D_4D_64} come from effective SNR gains of 4D-64PRS are 0.83 dB and 0.4 dB, which is relative to PM-8QAM and 4D-2A8PSK, resp.} These gains can be compared to those in Fig. \ref{fig:GMI_4D_64}, where no NLI is present (0.67 dB and 0.4 dB for PM-8QAM and 4D-2A8PSK respectively.) For the PM-8QAM case, an effective SNR gain is observed. The relative additional shaping gain in terms of GMI for the nonlinear channel with respect to the gain in the linear channel is $(0.27-0.24)/0.24=12.5\%$. For the 4D-2A8PSK case, the effective SNR remains the same, which comes from the fact that both 4D-64PRS and 4D-2A8PSK have the constant-modulus property. The gains of 4D-64PRS in the  nonlinear channel with respect to 4D-2A8PSK is almost the same as those in the AWGN channel ($\approx 0.4$~dB).

\subsection{Analysis of the 6~bit/4D-sym modulation formats}\label{sec:Discussion}

In this section, we analyze the structure of the proposed format (its coordinates and binary labeling are presented in Table~\ref{tab:4D_64_XL}) and give some intuition on why it outperforms other formats available in the literature. We first study the structure of the formats in terms of minimum squared Euclidean distance (MSED), which we denote by $d^2$. We will also look at the number of pairs of constellation points at MSED, which we denote as $n$. A large $d^2$ and small $n$ should in principle result in high MI in the high-SNR regime, as recently proved in \cite{AlvaradoTIT2014}\footnote{The results in \cite{AlvaradoTIT2014} hold for 1D constellations only. However, the authors in \cite{AlvaradoTIT2014} conjectured the results to hold verbatim to any number of dimensions.}.

From now on, we assume all the constellation are normalized to $E_s=2$ (i.e., unit energy per polarization). Under this assumption, the MSED of 4D-64PRS  is $d^2=0.69$ and {$n=32$}. On the other hand, 4D-2A8PSK has a MSED of $d^2=0.88$ and $n=128$, while 4D-64SP-12QAM has the highest MSED  $d^2=1$ but a rather large $n=272$. Based on these properties of the formats, 4D-64SP-12QAM and 4D-2A8PSK should be better than 4D-64PRS in terms of MI. Our numerical results, however, show that all these three formats have a very similar MI (see MI results in Fig.~\ref{fig:GMI_4D_64}). We believe that there are two reasons for this. Firstly, MSED might not be the only quantity to look at when moderate SNRs are considered, but instead, the whole SED ``spectrum'' should be studied. Secondly, a large number of pairs of constellation points at the MSED reduces the AIR. 4D-64SP-12QAM and 4D-2A8PSK have $8.5$ and $4$ times more pairs at MSED than 4D-64PRS, resp., which could explain the relatively low MI values obtained by those constellations. 

\begin{figure}[!tb]
\centering
   \includegraphics[width=\columnwidth]{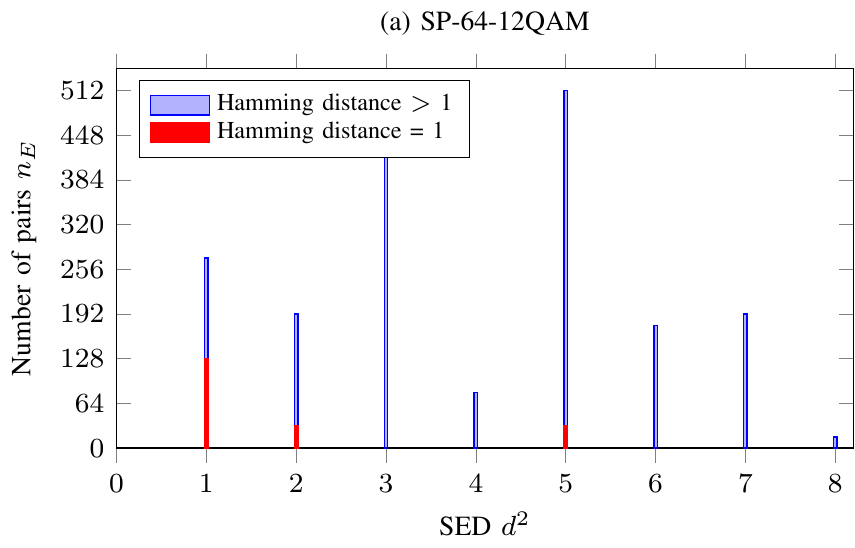}
   \includegraphics[width=\columnwidth]{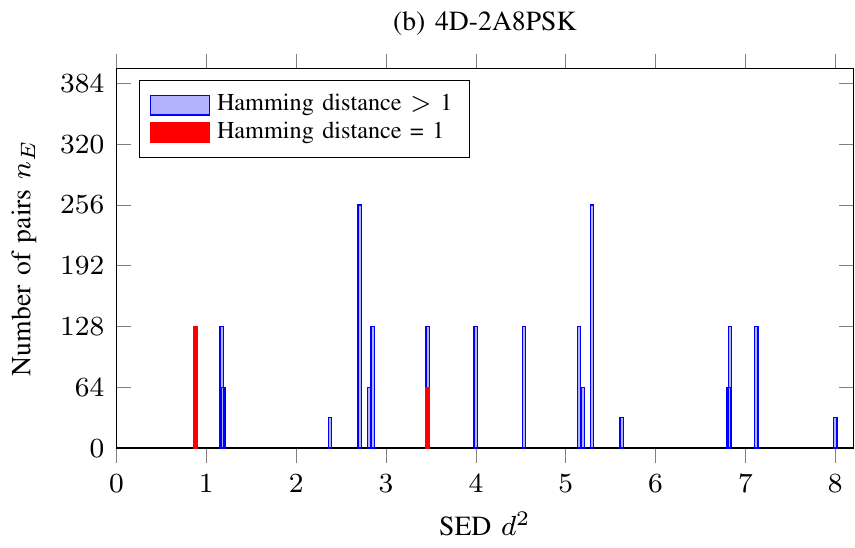}
   \includegraphics[width=\columnwidth]{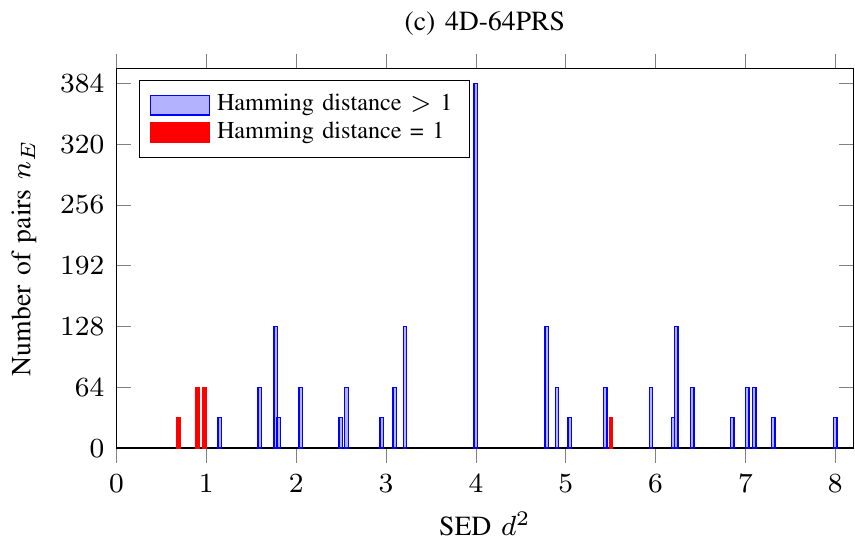}
  \caption{Histograms of SEDs of three 4D formats: (a) 4D-64SP-12QAM, (b) 4D-2A8PSK, and (c) 4D-64PRS. The red bars show the number of pairs with Hamming distance of 1 at the SED $d^2.$}
\label{fig:histSEDs}
\end{figure}

The first three columns of Table \ref{tab:compare} show a summary of these parameters, where we also include PM-8QAM (not analyzed in terms of MI because of the very low MI it provides). The spectrum of possible SEDs for the three constellations can also be seen in the x-axes of the histograms presented in Fig.~\ref{fig:histSEDs}.
 
 \begin{table}[!tb]
\caption{Comparison of 6 bit/4D-sym modulation formats}
\label{tab:compare}
\centering
\input{SEDTable.tex}
\end{table}

We now analyze GMI at medium  SNR  values. In this case, the quantities to investigate are not only MSED and the number of pairs at MSED, but also the Hamming distances (HDs) of the binary labels of the constellation points at MSED. The lower the Hamming distance the higher the GMI. In the extreme case where all the pairs of constellations points at MSED are at Hamming distance one, the binary labeling is said to be Gray. Interestingly, both 4D-2A8PSK and 4D-64PRS constellations fulfill this property, i.e., both constellations are Gray-labeled.\footnote{Note that a Gray-labeled 4D constellation is not necessarily Gray-labeled when considering the projection in two dimensions.} 

The discussion on the GMI performance is facilitated by the results in Fig. \ref{fig:histSEDs}, where histograms of the SEDs of the formats are shown. This figure also shows a classification of the pairs at a given SED: blue bars for pairs at HD larger than one, and red bars for pairs at HD one. This information is also shown in the last column of Table~\ref{tab:compare}, where the pairs $(d^2_{\text{HD}=1},n_{\text{HD}=1})$ are presented. These pairs show the number of pairs at HD one and the corresponding SEDs associated to those.

Fig.~\ref{fig:histSEDs} shows that 4D-64SP-12QAM has the highest MSED, however, it is worse in terms of GMI than 4D-2A8PSK and 4D-64PRS (see GMI results in Fig.~\ref{fig:GMI_4D_64}). This is due to the high number of pairs at MSED. We can observe that, even the MSED of 4D-64PRS is smaller than 4D-2A8PSK and 4D-64SP-12QAM, most of the points for 4D-64PRS with Hamming distance one have a larger SED than 4D-2A8PSK and PM-8QAM. Therefore, in terms of MSED, the performance of 4D-64PRS reduces, but it increases gain by reducing the number of pairs at MSED and also increases the SED for these points.

\section{Conclusions}\label{Sec:Conclusions}
This paper proposed and studied a new four-dimensional geometrically-shaped nonlinearity-tolerant modulation format with 6 bit/4D-sym named four-dimensional polarization-ring-switching (4D-64PRS). Based on the idea of imposing a constant-modulus on the 4D structure, the coordinates of the format and its binary labeling were \emph{jointly} optimized. The proposed format is optimal in terms of generalized mutual information. To the best of our knowledge, 4D-64PRS is the best performing nonlinearity-tolerant format proposed with this spectral efficiency.

{We believe that 4D-64PRS is a promising candidate for transmission systems with high nonlinearity.} In this paper, we concentrated on the AWGN channel and dispersion unmanaged links to show the superiority of the proposed 4D modulation format. A performance advantage on dispersion managed (strong nonlinearity) links is also expected and left for further investigation.

\balance
\bibliographystyle{IEEEtran}
\bibliography{references,reference_ECOC}

\end{document}

%% file: myFigures.tex
\usetikzlibrary{plotmarks,matrix,chains,scopes,fit,calc,shapes,positioning,decorations,intersections,arrows,backgrounds,shadows}

\newlength\FigureHeight
\newlength\FigureWidth


%% file: myStyles.tex
\definecolor{myDarkGreen}{rgb}{0.00000,0.58824,0.00000}%
\definecolor{uniform}{rgb}{0.00000,0.58824,0.00000}%
\definecolor{AWGNreference}{rgb}{0.00000,0.58824,0.00000}%

%% file: myMacros.tex
\DeclareMathOperator*{\argmax}{argmax}

\newcounter{lemma}

\newtheorem{exampleplain}{Example}

\newcommand{\bC}{\boldsymbol{C}}
\newcommand{\bb}{\boldsymbol{b}}
\newcommand{\bX}{\boldsymbol{X}}\newcommand{\bx}{\boldsymbol{x}}

\newcommand{\bY}{\boldsymbol{Y}}
\newcommand{\bZ}{\boldsymbol{Z}}

\newcommand{\bs}{\boldsymbol{s}}



%% file: myPackages.tex
\usepackage[utf8]{inputenc}

\usepackage{amsmath,amssymb}
\usepackage{xcolor}
\usepackage{graphicx}
\usepackage{balance}
\usepackage[pdftex]{hyperref} 
\hypersetup{colorlinks,linkcolor=black,citecolor=blue} 
\usepackage{cite}
\usepackage{multirow} 

\usepackage{gensymb}

%% file: SEDTable.tex
\scalebox{0.85}
{
\begin{tabular}{c|c|c|c|c}
\hline

\hline
{ } & { $d^2$} & { $n$} & {Gray-labeled } & ($d^2_{\text{HD}=1}$, $n_{\text{HD}=1}$)\\
\hline 

\hline
 \multirow{2}{*}{ PM-8QAM} & \multirow{2}{*}{0.84} &  \multirow{2}{*}{768}  &  \multirow{2}{*}{No} &  (0.84, 128)\\ &  & &  & (3.15, 64)    \\
 \hline
 \multirow{3}{*}{4D-64SP-12QAM} & \multirow{ 3}{*}{1} &  \multirow{3}{*}{272}  &  \multirow{ 3}{*}{No} &  (1, 128)\\ &  & &  & (2, 32) \\ &  & &  & (5, 32)\\
 \hline
\multirow{2}{*}{4D-2A8PSK} & \multirow{ 2}{*}{0.88} &  \multirow{2}{*}{128}  &  \multirow{ 2}{*}{Yes} &  (0.88, 128)\\ &  & &  & (3.46, 64)    \\
\hline
\multirow{ 4}{*}{4D-64PRS} & \multirow{ 4}{*}{0.69} &  \multirow{ 4}{*}{32}  &   \multirow{ 4}{*}{Yes} &  (0.69, 32)\\ &  & &  & (0.90, 64)\\ &  & &  & (0.98, 64)\\ &  & &  & (5.50, 32)     \\
\hline

\hline
\end{tabular}
}

%% file: ArxivV2.bbl
\begin{thebibliography}{10}
\providecommand{\url}[1]{#1}
\csname url@samestyle\endcsname
\providecommand{\newblock}{\relax}
\providecommand{\bibinfo}[2]{#2}
\providecommand{\BIBentrySTDinterwordspacing}{\spaceskip=0pt\relax}
\providecommand{\BIBentryALTinterwordstretchfactor}{4}
\providecommand{\BIBentryALTinterwordspacing}{\spaceskip=\fontdimen2\font plus
\BIBentryALTinterwordstretchfactor\fontdimen3\font minus
  \fontdimen4\font\relax}
\providecommand{\BIBforeignlanguage}[2]{{%
\expandafter\ifx\csname l@#1\endcsname\relax
\typeout{** WARNING: IEEEtran.bst: No hyphenation pattern has been}%
\typeout{** loaded for the language `#1'. Using the pattern for}%
\typeout{** the default language instead.}%
\else
\language=\csname l@#1\endcsname
\fi
#2}}
\providecommand{\BIBdecl}{\relax}
\BIBdecl

\bibitem{Buchali2016}
F.~Buchali, F.~Steiner, G.~Böcherer, L.~Schmalen, P.~Schulte, and W.~Idler,
  ``Rate adaptation and reach increase by probabilistically shaped 64-{QAM}: An
  experimental demonstration,'' \emph{Journal of Lightwave Technology},
  vol.~34, no.~7, pp. 1599--1609, April 2016.

\bibitem{TobiasJLT16}
T.~Fehenberger, A.~Alvarado, G.~B\"{o}cherer, and N.~Hanik, ``On probabilistic
  shaping of quadrature amplitude modulation for the nonlinear fiber channel,''
  \emph{Journal of Lightwave Technology}, vol.~34, no.~21, pp. 5063--5073, Nov
  2016.

\bibitem{BochererECOC2017}
G.~B\"{o}cherer, F.~Steiner, and P.~Schulte, ``Fast probabilistic shaping
  implementation for long-haul fiber-optic communication systems,'' in
  \emph{2017 European Conference on Optical Communication (ECOC)}, Sep. 2017,
  pp. 1--3.

\bibitem{Buchali2017}
F.~Buchali, W.~Idler, R.~Dischler, T.~Eriksson, and L.~Schmalen, ``Spectrally
  efficient probabilistically shaped square 64{QAM} to 256{QAM},'' in
  \emph{2017 European Conference on Optical Communication (ECOC)}, Sept 2017,
  pp. 1--3.

\bibitem{Maher2017}
R.~Maher, K.~Croussore, M.~Lauermann, R.~Going, X.~Xu, and J.~Rahn,
  ``Constellation shaped 66 {GB}d {DP}-1024{QAM} transceiver with 400 km
  transmission over standard {SMF},'' in \emph{2017 European Conference on
  Optical Communication (ECOC)}, Sept 2017, pp. 1--3.

\bibitem{trellisshaping}
G.~D. Forney, ``Trellis shaping,'' \emph{IEEE Transactions on Information
  Theory}, vol.~38, no.~2, pp. 281--300, March 1992.

\bibitem{lang1989}
G.~R. Lang and F.~M. Longstaff, ``A leech lattice modem,'' \emph{IEEE J. Sel.
  Areas Commun.}, vol.~7, no.~6, pp. 968--973, Aug 1989.

\bibitem{laroia1994}
R.~Laroia, N.~Farvardin, and S.~A. Tretter, ``On optimal shaping of
  multidimensional constellations,'' \emph{IEEE Transactions Information
  Theory}, vol.~40, no.~4, pp. 1044--1056, Jul 1994.

\bibitem{dyadic}
G.~B\"{o}cherer and R.~Mathar, ``Matching dyadic distributions to channels,''
  in \emph{2011 Data Compression Conference}, March 2011, pp. 23--32.

\bibitem{Qu2017}
Z.~Qu and I.~B. Djordjevic, ``Geometrically shaped 16{QAM} outperforming
  probabilistically shaped 16{QAM},'' in \emph{2017 European Conference on
  Optical Communication (ECOC)}, Sept 2017, pp. 1--3.

\bibitem{ZhangECOC2017}
S.~Zhang, F.~Yaman, E.~Mateo, T.~Inoue, K.~Nakamura, and Y.~Inada, ``Design and
  performance evaluation of a {GMI}-optimized 32{QAM},'' in \emph{2017 European
  Conference on Optical Communication (ECOC)}, Sept 2017, pp. 1--3.

\bibitem{Kojima2017JLT}
K.~Kojima, T.~Yoshida, T.~Koike-Akino, D.~S. Millar, K.~Parsons, M.~Pajovic,
  and V.~Arlunno, ``Nonlinearity-tolerant four-dimensional {2A8PSK} family for
  5-7 bits/symbol spectral efficiency,'' \emph{Journal Lightwave Technology},
  vol.~35, no.~8, pp. 1383--1391, April 2017.

\bibitem{Millar2018_OFC}
D.~S. Millar, T.~Fehenberger, T.~Koike-Akino, K.~Kojima, and K.~Parsons,
  ``Coded modulation for next-generation optical communications,'' in
  \emph{2018 Optical Fiber Communications Conference and Exposition (OFC)},
  March 2018, pp. 1--3.

\bibitem{BinECOC2018}
B.~Chen, C.~Okonkwo, H.~Hafermann, and A.~Alvarado, ``Increasing achievable
  information rates via geometric shaping,'' in \emph{2018 European Conference
  on Optical Communication (ECOC)}, Sept 2018, pp. 1--3.

\bibitem{BinICTON2018}
B.~Chen, C.~Okonkwo, D.~Lavery, and A.~Alvarado, ``Geometrically-shaped
  64-point constellations via achievable information rates,'' in \emph{2018
  20th International Conference on Transparent Optical Networks (ICTON)}, July
  2018, pp. 1--4.

\bibitem{8346117}
H.~G. Batshon, M.~V. Mazurczyk, J.~. Cai, O.~V. Sinkin, M.~Paskov, C.~R.
  Davidson, D.~Wang, M.~Bolshtyansky, and D.~Foursa, ``Coded modulation based
  on 56{APSK} with hybrid shaping for high spectral efficiency transmission,''
  in \emph{2017 European Conference on Optical Communication (ECOC)}, Sept
  2017, pp. 1--3.

\bibitem{Cai17OFC}
J.-X. Cai, H.~G. Batshon, M.~V. Mazurczyk, O.~V. Sinkin, D.~Wang, M.~Paskov,
  W.~Patterson, C.~R. Davidson, P.~Corbett, G.~Wolter, T.~Hammon,
  M.~Bolshtyansky, D.~Foursa, and A.~Pilipetskii, ``70.4 {Tb/s} capacity over
  7,600 km in {C+L} band using coded modulation with hybrid constellation
  shaping and nonlinearity compensation,'' in \emph{2017 Optical Fiber
  Communications Conference and Exhibition (OFC)}, Los Angeles, CA, Mar. 2017.

\bibitem{BICM_book}
L.~Szczecinski and A.~Alvarado, \emph{Bit-interleaved coded modulation:
  fundamentals, analysis and design}.\hskip 1em plus 0.5em minus 0.4em\relax
  John Wiley \& Sons, 2015.

\bibitem{Steiner2017}
F.~Steiner and G.~Boecherer, ``Comparison of geometric and probabilistic
  shaping with application to {ATSC} 3.0,'' in \emph{2017 International ITG
  Conference on Systems, Communications and Coding (SCC)}, Feb 2017, pp. 1--6.

\bibitem{SchulteWCL2019}
P.~{Schulte} and F.~{Steiner}, ``Divergence-optimal fixed-to-fixed length
  distribution matching with shell mapping,'' \emph{IEEE Wireless
  Communications Letters}, vol.~8, no.~2, pp. 620--623, April 2019.

\bibitem{GultekinISIT2018}
Y.~C. {G{\"u}ltekin}, F.~M.~J. {Willems}, W.~J. {van Houtum}, and
  S.~{Şerbetli}, ``Approximate enumerative sphere shaping,'' in \emph{2018
  IEEE International Symposium on Information Theory (ISIT)}, June 2018, pp.
  676--680.

\bibitem{GultekinarXiv2019}
\BIBentryALTinterwordspacing
Y.~C. {G{\"u}ltekin}, W.~J. {van Houtum}, A.~{Koppelaar}, and F.~M.~J.
  {Willems}, ``{Enumerative Sphere Shaping for Wireless Communications with
  Short Packets},'' \emph{arXiv e-prints}, Mar. 2019. [Online]. Available:
  \url{http://arxiv.org/abs/1903.10244}
\BIBentrySTDinterwordspacing

\bibitem{2018Tobias_PBDM}
T.~{Fehenberger}, D.~S. {Millar}, T.~{Koike-Akino}, K.~{Kojima}, and
  K.~{Parsons}, ``Multiset-partition distribution matching,'' \emph{IEEE
  Transactions on Communications}, vol.~67, no.~3, pp. 1885--1893, March 2019.

\bibitem{YoshidaECOC2016}
T.~{Yoshida}, K.~{Matsuda}, K.~{Kojima}, H.~{Miura}, K.~{Dohi}, M.~{Pajovic},
  T.~{Koike-Akino}, D.~S. {Millar}, K.~{Parsons}, and T.~{Sugihara},
  ``Hardware-efficient precise and flexible soft-demapping for
  multi-dimensional complementary {APSK} signals,'' in \emph{2016 European
  Conference on Optical Communication (ECOC)}, Sep. 2016, pp. 1--3.

\bibitem{BendimeradECOC2018}
D.~F. {Bendimerad}, H.~{Zhang}, and I.~{Land}, ``Ultra-low complexity and high
  performance soft-demapper for {8D} set-partitioned pdm-qpsk modulation
  formats,'' in \emph{2018 European Conference on Optical Communication
  (ECOC)}, Sep. 2018, pp. 1--3.

\bibitem{NakamuraJLT2018}
M.~{Nakamura}, F.~{Hamaoka}, A.~{Matsushita}, H.~{Yamazaki}, M.~{Nagatani},
  A.~{Hirano}, and Y.~{Miyamoto}, ``Low-complexity iterative soft-demapper for
  multidimensional modulation based on bitwise log likelihood ratio and its
  demonstration in high baud-rate transmission,'' \emph{Journal of Lightwave
  Technology}, vol.~36, no.~2, pp. 476--484, Jan 2018.

\bibitem{Shiner:14}
A.~D. Shiner, M.~Reimer, A.~Borowiec, S.~O. Gharan, J.~Gaudette, P.~Mehta,
  D.~Charlton, K.~Roberts, and M.~O'Sullivan, ``Demonstration of an
  8-dimensional modulation format with reduced inter-channel nonlinearities in
  a polarization multiplexed coherent system,'' \emph{Opt. Express}, vol.~22,
  no.~17, pp. 20\,366--20\,374, Aug 2014.

\bibitem{ReimerOFC2016}
M.~Reimer, S.~O. Gharan, A.~D. Shiner, and M.~O'Sullivan, ``Optimized 4 and 8
  dimensional modulation formats for variable capacity in optical networks,''
  in \emph{2016 Optical Fiber Communications Conference and Exhibition (OFC)},
  March 2016, pp. 1--3.

\bibitem{Bendimerad:18}
D.~F. Bendimerad, H.~Hafermann, and H.~Zhang, ``Nonlinearity-tolerant {8D}
  modulation formats by set-partitioning {PDM-QPSK},'' in \emph{2018 Optical
  Fiber Communications Conference and Exhibition (OFC)}, 2018, pp. 1--3.

\bibitem{Alvarado2015_JLT}
A.~Alvarado and E.~Agrell, ``Four-dimensional coded modulation with bit-wise
  decoders for future optical communications,'' \emph{Journal Lightwave
  Technology}, vol.~33, no.~10, pp. 1993--2003, May 2015.

\bibitem{Schmalen17}
L.~Schmalen, A.~Alvarado, and R.~Rios-M\"{u}ller, ``Performance prediction of
  nonbinary forward error correction in optical transmission experiments,''
  \emph{Journal of Lightwave Technology}, vol.~35, no.~4, pp. 1015--1026, Feb.
  2017.

\bibitem{AlvaradoJLT2015}
A.~Alvarado, E.~Agrell, D.~Lavery, R.~Maher, and P.~Bayvel, ``Replacing the
  soft-decision {FEC} limit paradigm in the design of optical communication
  systems,'' \emph{Journal of Lightwave Technology}, vol.~33, no.~20, pp.
  4338--4352, Oct 2015.

\bibitem{AlvaradoJLT2018}
A.~Alvarado, T.~Fehenberger, B.~Chen, and F.~M.~J. Willems, ``Achievable
  information rates for fiber optics: Applications and computations,''
  \emph{Journal of Lightwave Technology}, vol.~36, no.~2, pp. 424--439, Jan
  2018.

\bibitem{RasmusECOC2018}
R.~T. Jones, T.~A. Eriksson, M.~P. Yankov, and D.~Zibar, ``Deep learning of
  geometric constellation shaping including fiber nonlinearities,'' in
  \emph{2018 European Conference on Optical Communication (ECOC)}, Sept 2018,
  pp. 1--3.

\bibitem{MuellerOFC2015}
R.~Rios-Müller, J.~Renaudier, L.~Schmalen, and G.~Charlet, ``Joint coding rate
  and modulation format optimization for {8QAM} constellations using {BICM}
  mutual information,'' in \emph{2015 Optical Fiber Communications Conference
  and Exhibition (OFC)}, March 2015, pp. 1--3.

\bibitem{ZhangECOC2015}
S.~Zhang, K.~Nakamura, F.~Yaman, E.~Mateo, T.~Inoue, and Y.~Inada, ``Optimized
  {BICM-8QAM} formats based on generalized mutual information,'' in \emph{2015
  European Conference on Optical Communication (ECOC)}, Sept 2015, pp. 1--3.

\bibitem{NakamuraECOC2015}
T.~Nakamura, E.~L.~T. de~Gabory, H.~Noguchi, W.~Maeda, J.~Abe, and K.~Fukuchi,
  ``Long haul transmission of four-dimensional 64{SP}-12{QAM} signal based on
  16{QAM} constellation for longer distance at same spectral efficiency as
  {PM-8QAM},'' in \emph{2015 European Conference on Optical Communication
  (ECOC)}, Sept 2015, pp. 1--3.

\bibitem{DarISIT2014}
R.~Dar, M.~Feder, A.~Mecozzi, and M.~Shtaif, ``On shaping gain in the nonlinear
  fiber-optic channel,'' in \emph{2014 IEEE International Symposium on
  Information Theory}, June 2014, pp. 2794--2798.

\bibitem{Poggiolini_JLT2014}
P.~Poggiolini, G.~Bosco, A.~Carena, V.~Curri, Y.~Jiang, and F.~Forghieri, ``The
  {GN}-model of fiber non-linear propagation and its applications,''
  \emph{Journal of Lightwave Technology}, vol.~32, no.~4, pp. 694--721, Feb
  2014.

\bibitem{Dar:13}
R.~Dar, M.~Feder, A.~Mecozzi, and M.~Shtaif, ``Properties of nonlinear noise in
  long dispersion-uncompensated fiber links,'' \emph{Opt. Express}, vol.~21,
  no.~22, pp. 25\,685--25\,699, Nov 2013.

\bibitem{SecondiniPTL2012}
M.~Secondini and E.~Forestieri, ``Analytical fiber-optic channel model in the
  presence of cross-phase modulation,'' \emph{IEEE Photonics Technology
  Letters}, vol.~24, no.~22, pp. 2016--2019, Nov 2012.

\bibitem{SerenaECOC2013}
P.~Serena and A.~Bononi, ``On the accuracy of the {Gaussian} nonlinear model
  for dispersion-unmanaged coherent links,'' in \emph{2013 European Conference
  and Exhibition on Optical Communication (ECOC)}, Sep. 2013, pp. 1--3.

\bibitem{Carena:14}
A.~Carena, G.~Bosco, V.~Curri, Y.~Jiang, P.~Poggiolini, and F.~Forghieri,
  ``{EGN} model of non-linear fiber propagation,'' \emph{Opt. Express},
  vol.~22, no.~13, pp. 16\,335--16\,362, Jun 2014.

\bibitem{Dar:14}
R.~Dar, M.~Feder, A.~Mecozzi, and M.~Shtaif, ``Accumulation of nonlinear
  interference noise in fiber-optic systems,'' \emph{Opt. Express}, vol.~22,
  no.~12, pp. 14\,199--14\,211, Jun 2014.

\bibitem{PanJLT2016}
C.~Pan and F.~R. Kschischang, ``Probabilistic 16-{QAM} shaping in {WDM}
  systems,'' \emph{Journal of Lightwave Technology}, vol.~34, no.~18, pp.
  4285--4292, Sep. 2016.

\bibitem{YankovnJLT2016}
M.~P. Yankov, F.~D. Ros, E.~P. da~Silva, S.~Forchhammer, K.~J. Larsen, L.~K.
  Oxenløwe, M.~Galili, and D.~Zibar, ``Constellation shaping for {WDM} systems
  using {256QAM/1024QAM} with probabilistic optimization,'' \emph{Journal of
  Lightwave Technology}, vol.~34, no.~22, pp. 5146--5156, Nov 2016.

\bibitem{RennerJLT2017}
J.~Renner, T.~Fehenberger, M.~P. Yankov, F.~D. Ros, S.~Forchhammer,
  G.~Böcherer, and N.~Hanik, ``Experimental comparison of probabilistic
  shaping methods for unrepeated fiber transmission,'' \emph{Journal of
  Lightwave Technology}, vol.~35, no.~22, pp. 4871--4879, Nov 2017.

\bibitem{RasmusArxiv2018}
\BIBentryALTinterwordspacing
R.~T. Jones, T.~A. Eriksson, M.~P. Yankov, B.~J. Puttnam, G.~Rademacher, R.~S.
  Luis, and D.~Zibar, ``Geometric constellation shaping for fiber optic
  communication systems via end-to-end learning,'' \emph{arXiv}, 2018.
  [Online]. Available: \url{http://arxiv.org/abs/1810.00774}
\BIBentrySTDinterwordspacing

\bibitem{SillekensOFC2018}
E.~Sillekens, D.~Semrau, G.~Liga, N.~A. Shevchenko, Z.~Li, A.~Alvarado,
  P.~Bayvel, R.~I. Killey, and D.~Lavery, ``A simple nonlinearity-tailored
  probabilistic shaping distribution for square {QAM},'' in \emph{2018 Optical
  Fiber Communications Conference and Exposition (OFC)}, March 2018, pp. 1--3.

\bibitem{SillekensECOC2018}
E.~Sillekens, D.~Semrau, D.~Lavery, P.~Bayvel, and R.~I. Killey, ``Experimental
  demonstration of geometrically-shaped constellations tailored to the
  nonlinear fibre channel,'' in \emph{2018 European Conference and Exhibition
  on Optical Communication (ECOC)}, Sep. 2018, pp. 1--3.

\bibitem{DarJLT2015}
R.~Dar, M.~Feder, A.~Mecozzi, and M.~Shtaif, ``Inter-channel nonlinear
  interference noise in {WDM} systems: Modeling and mitigation,'' \emph{Journal
  of Lightwave Technology}, vol.~33, no.~5, pp. 1044--1053, March 2015.

\bibitem{Moore2009}
B.~Moore, G.~Takahara, and F.~Alajaji, ``Pairwise optimization of modulation
  constellations for non-uniform sources,'' \emph{Canadian Journal of
  Electrical and Computer Engineering}, vol.~34, no.~4, pp. 167--177, Fall
  2009.

\bibitem{ZhangOFC2017}
S.~{Zhang}, F.~{Yaman}, E.~{Mateo}, T.~{Inoue}, K.~{Nakamura}, and Y.~{Inada},
  ``A generalized pairwise optimization for designing multi-dimensional
  modulation formats,'' in \emph{2017 Optical Fiber Communications Conference
  and Exhibition (OFC)}, March 2017, pp. 1--3.

\bibitem{Schreckenbach2003}
F.~Schreckenbach, N.~Gortz, J.~Hagenauer, and G.~Bauch, ``Optimization of
  symbol mappings for bit-interleaved coded modulation with iterative
  decoding,'' \emph{IEEE Communications Letters}, vol.~7, no.~12, pp. 593--595,
  Dec 2003.

\bibitem{AlvaradoCommL2014}
A.~Alvarado, F.~Brannstrom, and E.~Agrell, ``A simple approximation for the
  bit-interleaved coded modulation capacity,'' \emph{IEEE Communications
  Letters}, vol.~18, no.~3, pp. 495--498, March 2014.

\bibitem{El-RahmanECOC2017}
A.~I.~A. El-Rahman and J.~C. Cartledge, ``Multidimensional geometric shaping
  for {QAM} constellations,'' in \emph{2017 European Conference on Optical
  Communication (ECOC)}, Sep. 2017, pp. 1--3.

\bibitem{ShenLIECOC2018}
S.~Li, C.~H\"{a}ger, N.~Garcia, and H.~Wymeersch, ``Achievable information
  rates for nonlinear fiber communication via end-to-end autoencoder
  learning,'' in \emph{2018 European Conference on Optical Communication
  (ECOC)}, Sep. 2018, pp. 1--3.

\bibitem{RenaudierECOC2012}
J.~Renaudier, A.~Voicila, O.~Bertran-Pardo, O.~Rival, M.~Karlsson, G.~Charlet,
  and S.~Bigo, ``Comparison of set-partitioned two-polarization {16QAM} formats
  with {PDM-QPSK} and {PDM-8QAM} for optical transmission systems with
  error-correction coding,'' in \emph{2012 European Conference and Exhibition
  on Optical Communications (ECOC)}, Sept 2012, pp. 1--3.

\bibitem{AlvaradoTIT2014}
A.~Alvarado, F.~Br\"{a}nnstr\"{o}m, E.~Agrell, and T.~Koch, ``High-{SNR}
  asymptotics of mutual information for discrete constellations with
  applications to {BICM},'' \emph{IEEE Transactions on Information Theory},
  vol.~60, no.~2, pp. 1061--1076, Feb 2014.

\end{thebibliography}
